\documentclass[journal]{IEEEtran}
\ifCLASSINFOpdf
\else
\fi

\usepackage{graphicx}
\usepackage{amssymb}
\usepackage{array}
\usepackage[flushleft]{threeparttable}
\usepackage{multirow}
\usepackage{rotating}
\usepackage{amsmath}
\usepackage{longtable}
\usepackage{pdflscape}
\usepackage{array}
\usepackage [english]{babel}
\usepackage [autostyle, english = american]{csquotes}
\usepackage{mathtools}
\usepackage{tabularx}
\usepackage{xcolor}
\usepackage{colortbl}
\usepackage{lipsum}
\usepackage[autostyle]{csquotes}
\usepackage{array,booktabs,ragged2e}
\usepackage{booktabs}
\usepackage{color}
\usepackage{float}
\usepackage{cancel}
\usepackage{dblfloatfix} 
\usepackage[normalem]{ulem}
\newcommand\redsout{\bgroup\markoverwith{\textcolor{red}{\rule[0.5ex]{2pt}{0.4pt}}}\ULon}
\newcommand{\sob}[1]{\textcolor{black}{#1}}
\newcommand{\sobia}[1]{\textcolor{black}{#1}}

\usepackage{hyperref}
\begin{document}
\title{\sobia{Information-Centric Networking based Internet of Things: A Survey and Future Research Directions}} 
\title{\sobia{Recent Advances in Information-Centric Networking based Internet of Things (ICN-IoT)}} 
\author{Sobia~Arshad, Muhammad~Awais~Azam, Mubashir Husain~Rehmani, Jonathan~Loo
\thanks{Sobia Arshad and Muhammad Awais Azam are with the Department of Computer Engineering, University of Engineering \& Technology, Taxila, Pakistan. (email(s):sobia.arshad@uettaxila.edu.pk, awais.azam@uettaxila.edu.pk) Phone:+92 (0)331 9811899}
\thanks{Mubashir Husain Rehmani is with the Waterford Institute of Technology, Ireland.(email: mshrehmani@gmail.com) Phone:+92 (0)333 3052764}
\thanks{Jonathan Loo is with the University of West London, London, UK. (jonathan.loo@uwl.ac.uk)}
\thanks{"Copyright (c) 2012 IEEE. Personal use of this material is permitted. However, permission to use this material for any other purposes must be obtained from the IEEE by sending a request to pubs-permissions@ieee.org."}
\thanks{Manuscript received Month March, 2018; revised Month July, 2018; accepted Month July, 2018.}}
\markboth{IEEE Internet of Things Journal,~Vol.~14, No.~8, September~2018}%
{Shell \MakeLowercase{\textit{et al.}}: Bare Demo of IEEEtran.cls for IEEE Journals}

\maketitle
\begin{abstract}
Information-Centric Networking (ICN) is being realized as a promising approach to accomplish the shortcomings of current IP-address based networking. ICN models are based on naming the content to get rid of address-space scarcity, accessing the content via name-based-routing, caching the content at intermediate nodes to provide reliable, efficient data delivery and self-certifying contents to ensure better security. Obvious benefits of ICN in terms of fast and efficient data delivery and improved reliability raises ICN as highly promising networking model for Internet of Things (IoTs) like environments. IoT aims to connect anyone and/or anything at any time by any path on any place. From last decade, IoTs attracts both industry and research communities. IoTs is an emerging research field and still in its infancy. 
 Thus, this paper presents the potential of ICN for IoTs by providing state-of-the-art literature survey. 
We discuss briefly the feasibility of ICN features and their models (and architectures) in the context of IoT. Subsequently, we present a comprehensive survey on ICN based caching, naming, security and mobility approaches for IoTs with appropriate classification. Furthermore, we present operating systems (OS) and simulation tools for ICN-IoT. Finally, we provide important research challenges and issues faced by ICN for IoTs.
\end{abstract}


\IEEEpeerreviewmaketitle
\section{Introduction}
\subsection{Motivation and Background}
IoTs aim to connect each and every device with the Internet, so that these devices can be accessed at any time, at any place and by any path (i.e., from any network) \cite{iot2}. IoTs canopies enchanted objects like smart washing machines, smart refrigerators, smart microwave ovens, smart-phones, smart meters and smart vehicles. Connectivity of these smart objects with the Internet enables many valuable and remarkable applications like smart home, smart building, smart transport, digital health, smart grid and smart cities. When billions of these devices connect to the Internet, generation of large amount of data is an apparent consequence. Moreover, this IoT data has to combine with the data produced from Facebook likes and YouTube videos which results in IoT Big Data. Therefore, efficient access and discovery of IoT Big Data put more constraints on the underlying TCP/IP architecture while raising many important issues. 

Among these issues (from the IoT device perspective), one is naming (and addressing) every IoT device \cite{atzori2010internet}-\cite{gubbi2013internet}. As IPv4 addressing space is exhausted, IPv6 address space may also exhaust in the future. Besides this, IPv6 address is quite long and its long length makes it less suitable for communication through constraint-oriented devices like wireless sensors \cite{shang2016challenges}-\cite{Borgia2014}-\cite{Stankovic2014}. Therefore, efficient naming and addressing schemes for billions of devices (and contents) are not ideally available in IP-architecture. Furthermore, every device has different constraints and specifications which raise another issue of heterogeneity. This is due to the fact that IoTs comprises on devices which are heterogeneous in terms of processing power capability, size, memory, battery life and cost. Moreover, most of the devices are tiny, low power, limited memory, low cost and constraint-oriented wireless sensors. These devices are usually known as smart devices. Besides heterogeneity, in these low memory and low battery life constraint-oriented devices, data can become unavailable most of the time which causes data unavailability. Therefore, solutions like in-network caching (which are required to make data available) are missing in naive IP-based networking. In addition, IoTs applications like smart home, smart town, smart grid and smart health requires more security and extra privacy in terms of data accessed by these devices and their usage \cite{varadharajan2016data}. Moreover, some IoTs applications, for instance, VANETs, MANETs and smart transport require better mobility handling \cite{SilvaSilvaBoavida2015}-\cite{Al-NidawiYahyaKemp2015}. 
       
On the other hand, from data perspective, most of the IoTs application users are more interested in getting the updated information rather than knowing the address of information source. As an instance, IoT devices especially in the domain called wireless sensor networks (WSN), have specific purpose to harvest information at the large scale \cite{Al-FuqahaGuizaniMohammadiEtAl2015}. Every device has to perform some specific task, for example temperature sensors measure temperature from their surroundings and does not perform word processing task that a general purpose computer does. Any user of temperature measurement application is interested in current temperature value of a certain area rather than the temperature value from a specific sensor.  
              
       \begin{table*}[!t]
\renewcommand{\arraystretch}{1.2}
\caption{List of Acronyms Used}
\label{table_acronyms}
\begin{tabularx}{\textwidth}{|m{1.5cm}|m{6.35cm}|m{1.5cm}|m{7cm}|}

\hline
\toprule
\bfseries Acronyms & \bfseries Definitions & \bfseries Acronyms & \bfseries Definitions\\
\hline
\toprule
\bfseries 6LowPANs & \textit{IPv6 over Low power Wireless Personal Area Networks} &
 
\bfseries CCN & \textit{Content-Centric Networking}\\
\hline
\bfseries CS & \textit{Content Store} &
 

\bfseries COMET & \textit{COntent Mediator architecture for content aware nETworks} \\ \hline
\bfseries CONET & \textit{Content Network}&

\bfseries DF & \textit{Destination Flag} \\ \hline


\bfseries DONA & \textit{Data Oriented Network Architecture} &
\bfseries DoS attack & \textit{Denial-of-Service attack} \\ \hline

\bfseries DPI  & \textit{Deep Packet Inspection} &

\bfseries FIA & \textit{Future Internet Architecture} \\ \hline

\bfseries FIA-NP & \textit{FIA-Next Phase} &
\bfseries FIB & \textit{Forwarding Information Base} \\ \hline

\bfseries FP7 & \textit{Framework Programme 7}&


\bfseries GPRS & \textit{General Packet Radio Service} \\ \hline

\bfseries GSM & \textit{Global System for Mobile communication} &

\bfseries GUID & \textit{Globally Unique Identifier}\\ \hline



\bfseries IERC & \textit{IoT European Research Cluster} &

\bfseries ICN & \textit{Information Centric Networking} \\
\hline
\bfseries IoT & \textit{Internet of Things}&
\bfseries  IPV4 & \textit{Internet Protocol version 4}\\
\hline
\bfseries IPv6 & \textit{Internet Protocol version 6}&
\bfseries LRU  & \textit{Least Recently Used} \\ \hline
\bfseries LTE & \textit{Long Term Evolution} &
\bfseries LTE-A & \textit{LTE Advanced} \\ \hline

\bfseries M2M & \textit{Machine-to-Machine} &
\bfseries MF & \textit{MobilityFirst}\\
\hline
\bfseries NDN & \textit{Named Data Networking}  &
\bfseries NetInf & \textit{Networking of Information}\\
\hline
\bfseries NFC & \textit{Near Field Communication}&
\bfseries NRS & \textit{Name Resolution System}\\
\hline
\bfseries NSF & \textit{National Science Foundation}&


\bfseries PARC & \textit{Palo Alto Research Center} \\ \hline
\bfseries PIT & \textit{Pending Interest Table} &

\bfseries PSIRP & \textit{Publish-Subscribe Internet Routing Paradigm}\\ \hline

\bfseries PURSUIT & \textit{Publish SUbscribe Internet Technology}&






\bfseries SAIL & \textit{Scalable and Adaptive Internet soLutions}\\ \hline

\bfseries SIT & \textit{Satisfied Interest Table} &



\bfseries TCP/IP & \textit{Transmission Control Protocol/Internet Protocol}\\ \hline







\hline



\end{tabularx}
\end{table*}
 Considering TCP/IP as network architecture for IoTs, which was traditionally designed to connect limited number of computers and to share limited and expensive network resources through limited address space at network layer, it is definitely not designed to fulfil IoTs requirements. Moreover, besides above-mentioned requirements, IoTs huge data put additional requirements like data dissemination and scalability on the underlying architecture. To fulfil all these needs of IoTs, Information-Centric Networking (ICN) (which is a promising candidate for the future Internet foundation) has recently emerged as an ideal candidate. So far, there are nine major architectures proposed under the concept of ICN including DONA, CCN \cite{ndnProject}, PURSUIT \cite{pursuit}, NetInf \cite{netinf}, CURLING \cite{comet-curling}, CONET \cite{convergence}, MobilityFirst \cite{mobilityfirst}, C-DAX \cite{powergridICN} and Green ICN \cite{greenICN}. Among these ICN-based architectures DONA, SAIL, COMET and CONVERGENCE, CCN all are dirty-slate while MF, PURSUIT and NDN are clean-slate architectures. 
        CCN (NDN) is prevailing approach among other ICN-based proposed architectures \cite{ccnx-prominent-proof-of-concept}. ICN primary characteristics include in-network caching, naming the contents, better and easy mobility management, improved security and scalable information delivery which are naturally suitable for IoT applications. Moreover, ICN-based hourglass architecture provides us thin-waist like TCP/IP \cite{ICNsurvey2012}. Additionally, ICN can mask over TCP/IP network layer or MAC layer. CCN could be applied just above MAC layer especially in WSN. 
        Current literature \cite{ICNsurvey}-\cite{AmadeoCampoloIeraEtAl2015} argue that ICN seems to replace IP; rather we believe and foresee ICN is an overlay network sitting on IP network. In fact, CCN is a layer that masks the need of associating content with the IP address instead by name. The actual content delivery still requires TCP/IP interface or direct MAC (layer 2) interface. 
       
       ICN's striking feature in-network caching, can efficiently handle the issue of information delivery from dead (unavailable) device due to low battery life by caching contents at intermediate nodes. Also it can minimize retrieval delay even in case of alive devices through the use of caching. Furthermore, naming the contents can resolve the address space scarcity issue of IPv4 and can enable scalability in an efficient way and can also offer better name management and easy information retrieval of huge data produced by IoT applications. 
       Moreover, mobility handling provides better hand-off for mobile devices like mobile phones and vehicles. ICN's self-certifying contents provide more security to data rather than securing the hosts \cite{FotiouPolyzos2014a}-\cite{ICNsurvey}. \sobia{Therefore, these are the reasons that in this article we survey ICN-based naming, in-network caching, security and mobility schemes which are explored for IoTs.} List of acronyms used in this paper is provided in Table~\ref{table_acronyms}.

\begin{table*}[!t]
\renewcommand{\arraystretch}{1.2}
\caption{IoTs and ICN Related Survey Articles}
\label{table_IoT_ICN_Surveys}
\centering
\begin{tabular}{c|c|l|c|c}
\hline
\toprule
\multicolumn{4}{|c|}{IoTs Related Survey Articles} \\
\hline
\toprule
\bfseries Sr\# &  \bfseries Reference(s) & \bfseries Topics Covered & \bfseries Publication Year\\
\midrule
1. 	& \cite{RazzaqueMilojevic-JevricPaladeEtAl2016}	& IoT middleware requirements and solutions & 2016 \\
\hline
2. & \cite{GranjalMonteiroSaSilva2015}& \begin{tabular}[c]{@{}l@{}}IoT security Issues and their corresponding solutions\\ \end{tabular} & 2015 \\
\hline
3. & \cite{HahmBaccelliPetersenEtAl2015} & Classification of IoT Oss & 2015 \\ 
\hline
4. &  \cite{Stankovic2014}&  Eight research directions for IoTs & 2014\\
\hline
5. &  \cite{PereraZaslavskyChristenEtAl2014}	& Context awareness solutions for IoT & 2014	\\
\hline
6. & \cite{ZhangLiangLuEtAl2014} & \begin{tabular}[c]{@{}l@{}}Sybil attacks in IoTs have been discussed along with defense schemes\\ \end{tabular} & 2014 \\ \hline
7. &\cite{Al-FuqahaGuizaniMohammadiEtAl2015}  & \multirow{3}{*}{\begin{tabular}[c]{@{}l@{}}Basics of IoT including building blocks and characteristics of IoTs, IoT \\enabling technologies, smart potential  applications,\\ projects and related research challenges\end{tabular}} & 2015 \\ \cline{1-2} \cline{4-4} 
8. & \cite{Borgia2014} &  & 2014 \\ \cline{1-2} \cline{4-4} 
9. & \cite{atzori2010internet} &  & 2010 \\ 
\hline
\toprule
\multicolumn{4}{|c|}{ICN Related Survey Articles} \\
\hline
\toprule
\bfseries Sr\# &  \bfseries Reference(s) & \bfseries Topics Covered & \bfseries Publication Year\\
\midrule
1. & \cite{AmadeoCampoloMolinaro2016}& ICN for VANETs and Future Directions &	2016 \\
\hline
2. 	& \cite{AbdAllahHassaneinZulkernine2015}	& Taxonomy of security attacks and naming corresponding solutions &	2015 \\
\hline
3. &  \cite{ZhangLuoZhang2015}	& caching mechanisms, performance parameters &	2015	\\
\hline
4. &  \cite{FangYuHuangEtAl2014}& {\begin{tabular}[c]{@{}l@{}} ICN energy efficient caching schemes, content placement, cache\\  placement and request-to-cache routing\end{tabular}} &	2014\\
\hline
5. &  \cite{ICNsurvey}	& Seven ICN Architectures and Research Directions & 2014\\
\hline
6. &	 \cite{BariChowdhuryAhmedEtAl2012}	& Routing and naming schemes &	2012\\
\hline
7. & \cite{ICNsurvey2012} & Four ICN Architectures & 2012\\
\hline

\toprule
\multicolumn{4}{|c|}{ICN for IoT Survey Article} \\
\hline
\toprule
\bfseries Sr\# &  \bfseries Reference & \bfseries Topics Covered & \bfseries Publication Year\\
\midrule
1. & \cite{AmadeoCampoloQuevedoEtAl2016}& Briefly identify ICN for IoT and Future Directions &	2016 \\
\bottomrule
\end{tabular}
\end{table*}



\subsection{Review of Related Survey Articles}
\sobia{Our current survey on ICN-based IoTs is unique from the prior surveys as we survey holistically ICN-based IoTs caching, ICN-based IoTs naming, ICN-based IoTs security and ICN-based IoTs mobility schemes.} A plenty of surveys is available on either alone IoTs or on specifically ICN related issues. To the best of our knowledge, this work is the only detailed survey that emphasizes ICN for IoTs.   
 
Exclusively IoT emphasized surveys have covered the IoT basics including building blocks and characteristics, enabling technologies, smart potential applications, projects and related research challenges in \cite{atzori2010internet}, \cite{Borgia2014}, \cite{Al-FuqahaGuizaniMohammadiEtAl2015}. Different eight research directions for IoTs are listed down in \cite{Stankovic2014}. Context awareness solutions for IoTs are discussed in \cite{PereraZaslavskyChristenEtAl2014}. Middle-ware requirements and solutions are surveyed in \cite{RazzaqueMilojevic-JevricPaladeEtAl2016}. IoTs security issues and their corresponding solutions are outlined in \cite{GranjalMonteiroSaSilva2015}. In \cite{ZhangLiangLuEtAl2014}, specifically Sybil attacks in IoTs are discussed along with their defense schemes. Moreover, classification of Operating Systems (OSs) for IoTs is presented in \cite{HahmBaccelliPetersenEtAl2015}. List of survey paper for IoTs is provided in Table~\ref{table_IoT_ICN_Surveys}. 

Surveys that solely focused ICN include \cite{ICNsurvey2012}, in which general ICN is described along with four ICN architectures including DONA, CCN, PSIRP and NetInf. George Xylomenos et al., in \cite{ICNsurvey} described ICN concept, its features and extended the research of \cite{ICNsurvey2012} by adding three more updated architectures named CONVERGENCE, CONET and MobilityFirst. Moreover, \cite{FangYuHuangEtAl2014} focused on ICN energy efficient caching schemes on the basis of content placement, cache placement and request-to-cache routing. While \cite{ZhangLuoZhang2015} discussed only NDN and DONA architectures, summarized caching mechanisms, described performance parameters and conducted simulations for the evaluation of caching mechanisms. Routing and naming schemes for ICN are covered in \cite{BariChowdhuryAhmedEtAl2012}. Comprehensive survey of possible attacks in ICN is presented in \cite{AbdAllahHassaneinZulkernine2015}. Moreover, taxonomy of security attacks (i.e., categorized into naming, caching, routing and other attacks) in ICN is presented and their existing solutions are discussed. ICN for VANETs along with future research directions is presented in \cite{AmadeoCampoloMolinaro2016}. ICN related literature survey is listed in Table~\ref{table_IoT_ICN_Surveys}.

One pioneer short article \cite{AmadeoCampoloQuevedoEtAl2016} that identifies ICN for IoT, surveys briefly ICN for IoT without providing enough literature survey details. \sobia{In contrast to \cite{AmadeoCampoloQuevedoEtAl2016}, our present survey, provides comprehensive up-to-date review of ICN for IoT, including ICN models and their feasibility for IoT, additionally caching techniques, naming schemes, security schemes and mobility handling mechanisms along with operating systems, simulators and detail research challenges for ICN-IoT research community.}

\subsection{Contribution of This Survey Article}
We mainly aim to discuss ICN for IoTs. To meet our aim, we provide holistic and comprehensive literature on ICN-based in-network caching, ICN content naming schemes, ICN security schemes and ICN mobility handling schemes for IoTs. 
With such goals, to the best of our knowledge, it makes this paper pioneer and unique in this field. We make the following contributions in this paper:
\begin{itemize}
\item We provide very brief overview of IoT architecture requirements and major ICN architectures with respect to their suitability for IoTs in terms of naming, caching, security and mobility handling schemes.
\item We summarize ICN-based architectures for IoT.
\item \sobia{We provide comprehensive survey of ICN-based in-network caching techniques for IoTs and classification of these schemes on the basis of role of content and node properties in ICN caching mechanisms for IoT.}
\item \sobia{We provide classification of ICN-based content naming approaches on the basis of name structures for IoTs.}
\item \sobia{We classify ICN-based security schemes for IoTs on the basis of their security handling for IoT contents and IoT devices.}
\item \sobia{We categorize ICN-based mobility schemes into IoT producer mobility and hand-off management.}
\item \sobia{We classify famous ICN-IoT simulators and OSs and identify ndnSIM as a more explored tool for ICN-IoT.} 
\item We provide issues, challenges and future research directions which ICN is facing for IoTs.
\end{itemize}
\subsection{Organization of the paper}
The rest of the paper is organized as follows. Section II provides a brief overview about IoT network architecture requirements, ICN models feasibility for IoT with respect to their naming, caching, security and mobility handling mechanisms and ICN-based architectures for IoTs. In sections III, IV, V, VI, ICN-based caching techniques, naming approaches, security and mobility support are discussed, respectively.
Section VII presents available OSs and simulators for ICN-IoTs. In section VIII, we present open challenges and future trends of ICN into IoT. Finally, section IX concludes the paper.
\section{Information-Centric Networking (ICN) Suitability for IoTs}
\begin{figure}[!t]
\centering
\includegraphics[width=10cm]{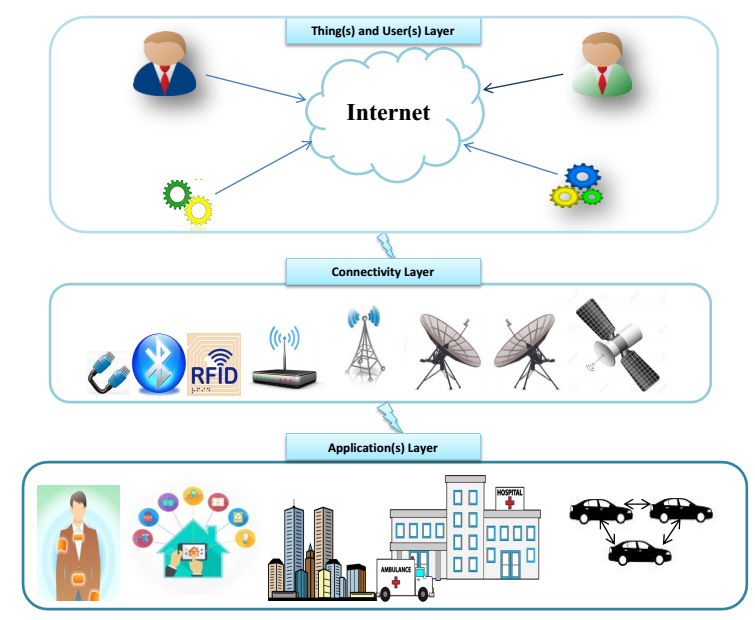}
\caption{Internet of Things (IoTs): Connectivity Types, Internet Technologies and IoTs Smart Applications.}
\label{iot}
\end{figure}

As IoTs is the connectivity of things through the unified Internet. Therefore, things can be humans and smart machines of any sort and this is illustrated in the lower portion of Fig.~\ref{iot}. These things can connect in three ways (connectivity in IoTs can be seen in upper portion of Fig.~\ref{iot}): i) Machine-Type-Communication (MTC), ii) Machine-to-Human (M2H) and iii) Human-to-Human (H2H). IoT works in four major steps namely:  i) Data acquisition or data sensing, ii) Data transmission,  iii) Data Processing and Information management and iv) Action \& Utilization. These major IoT working phases and corresponding elements can be visualized in Fig.~\ref{iot_Life_Cycle} and related literature is listed in Table.~\ref{table-IoT-LifeCycle-References}. 

This section fulfils six purposes: Firstly, we list and describe IoTs architecture requirements. However, our aim is not to survey and discuss IoTs in depth rather we illustrate it to highlight the related issues and identify architecture requirements. Secondly, we discuss IP-based evolutionary approaches for IoTs. Thirdly, we present the limitations of IP-based approaches. Fourthly, we provide mapping of IoT requirements against ICN characteristics. In next sub-section, we describe briefly ICN-based proposed architectures with respect to their naming, caching, security and mobility feasibility for IoTs and lastly, we present some approaches which discuss and explore ICN for IoTs.
\subsection{IoTs Architecture Requirements}
Specific requirements and challenges \cite{atzori2010internet, Al-FuqahaGuizaniMohammadiEtAl2015,Borgia2014} introduced by IoT network architecture outlined and given below:
\subsubsection{Scalability}
As IoTs envisions not only connecting networks and corresponding devices but enabling low power devices in billions to connect through the Internet. Thus, it imposes new challenges over underlying architecture in terms of scalability. IoTs architecture needs to support billions of devices in an efficient way. A current solution like IPv6 has a huge address space which can serve IoT devices. Although in future, addressing the IoT devices is not the only issue. Another case is a large amount of data which is being produced by IoT devices, also needs better and efficient scalability management. Therefore, IoTs network architecture must be explored in terms of scalability and it should be scalable to content access with network efficiency.
\subsubsection{Mobility} 
Mobile devices like tablets, smart-phones have a small screen and limited battery life. Moreover, some IoTs applications involve and require anytime and anywhere connectivity, in which users want to check their emails and/or make calls at anywhere and anytime. Furthermore, the number of mobile devices connecting to the Internet exceeds the stationary nodes. Therefore, to make data available at everywhere and provide fast and reliable connectivity, network architecture should support seamless mobility and roaming. 
\subsubsection{Security and Privacy} In some IoT scenarios like smart health and smart hospital; data that needs to be transmitted is highly sensitive. If any hacker tries to change it, it can lead to the alarming condition. To enable IoT efficiently, it should provide authorization, confidentiality, and integrity. Standards are needed to specify the data access policies like who can access the data and who cannot. Take the example of a smart home where the detail of a pizza ordered by the house owner is required by a pizza shop to charge the payment. If this detail is shared with his doctor or insurance company, this can affect user privacy. As insurance company is not the tentative user and could use the private data in the wrong way. Therefore, privacy must be ensured via some access policies.
\subsubsection{Naming and Addressing} IoT consists of billions of tiny, low-power and constraint-oriented devices which need unique names or addresses to get recognition in the IoT network. For example, if we talk about a single nano-network which may contain thousands of nano-nodes and then interconnection of many nano-networks would require complex IDs or addresses. Although large address space is available in IPv6, it may help in addressing and naming problem of IoT devices. But for constraint-oriented simple devices, it would be complex to process long address for very small communication because it results in the wastage of resources. On the other hand, IoTs contents are being produced and processed at very fast speed. In addition, there can be many versions or values against any single content with different time stamps. Naming these rapidly produced IoT contents adds further complications. Thus, a larger and permanent naming scheme and addressing space are still highly needed for both IoTs contents and devices.
\subsubsection{Heterogeneity and Interoperability} It can be seen above that RFID tags and smart wireless sensors mainly build IoTs. Smart sensors being major components of IoTs offer many applications. These devices are heterogeneous in nature and usually vary in specifications like memory size, processing power, and battery life. Moreover, communication between these sensors is carried out by different underlying technologies (wired, wireless, cellular, Bluetooth, 4G, LTE, CRN, opportunistic networks). Thus, heterogeneous technologies are involved in communication. Therefore, IoT network architecture is required to support heterogeneity among device specifications and different underlying communication technologies and techniques in an interoperable way.   
\subsubsection{Data Availability} In the current TCP/IP-based architecture, whenever a node moves from one location to another, data which is assumed to provide, becomes unavailable. The same case also occurs when some device runs out of battery and is not capable to forward data. In addition, Internet users cannot receive data at a time due to an occurrence of denial of service (DoS) attack. DoS occurs because the current Internet architecture cannot look at or inspect data according to request during data transmission. Consequently, methods like in-network caching are required to make data available with absolute certainty. 
\subsubsection{Energy Efficiency} Obviously billions of devices would need the huge amount of energy to build IoTs applications. Moreover, most of the smart devices are low in battery life such as wireless sensors. Thus, energy efficient mechanisms are required to make this universal connectivity possible in the form of IoTs.
\begin{figure}[!t]
\centering
\includegraphics[width=9cm]{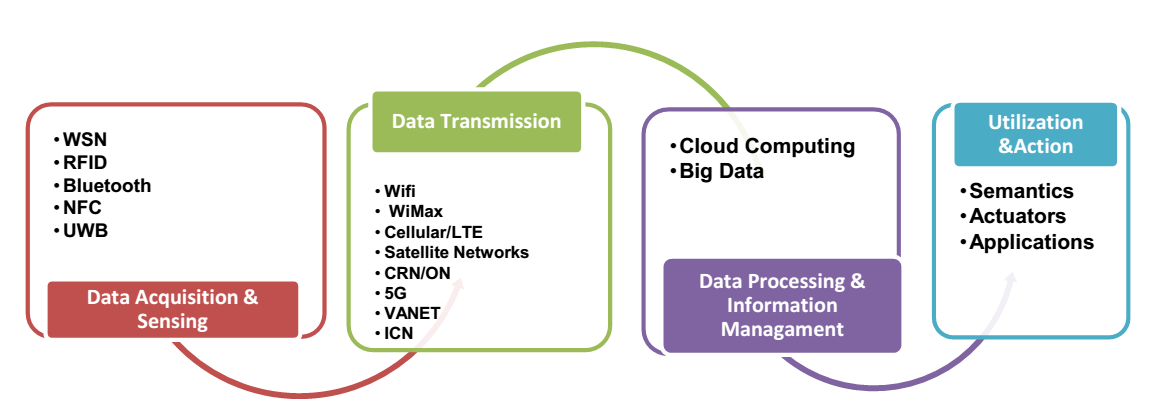}
\caption{Phases in IoT and Corresponding Enabling Technologies}
\label{iot_Life_Cycle}
\end{figure} 
\subsection{Evolutionary TCP/IP Approaches for IoTs}
To attain the above-mentioned requirements and recent trends about IoT architecture, these have prompted many research organizations to initiate multiple projects. Therefore, many evolutionary (or dirty slate) approaches are being explored for IoTs, for instance, IPv6-based 6LoWPANs \cite{iot-ipv6-project}-\cite{iot-ipv6}-\cite{iot-ipv6-2}.

Among these approaches, most of the projects are working under Internet Engineering Task Force (IETF). IETF projects are designing protocols for constraint-oriented devices based networks. 
The Constrained RESTful Environments (CoRE) \cite{core} group designed a framework for smart applications to work efficiently on IPv6-based constraint-oriented smart devices.  The Constrained Application Protocol (CoAP) \cite{coap} is a significant achievement that accomplished under CoRE working group. CoAP is a lighter version of the HTTP protocol. CoAP is designed mainly for low power devices forming constrained networks. CoAP also supports various caching forms which are mentioned in the REpresentational State Transfer (REST) protocol. CoAP runs over UDP to provide better communication among resource-oriented devices.

The IPv6 over Low Power Wireless Personal Area Networks working group (6LoWPANWG) \cite{6lowpans} has focused on 6LoWPANs. This group works for adaption of IPv6 over IEEE 802.15.4-based networks. The 6LoWPAN group also works for IPv6 header compression to efficiently run over low power devices. 

The Routing Over Low power and Lossy networks working group (ROLL) \cite{roll} mainly focus on developing routing strategies and self-configurable mechanisms in low power networks. Low power and Lossy networks (LLN) made up of many embedded devices which include limited power and memory devices. LLN provides an
end to end IP-based solution for routing over these networks. 6LoWPAN-WG will work closely to ROLL. Sometimes situations can happen in IoT when constraint-oriented devices are required to communicate with each other without any gateway. Therefore, IETF has designed the IPv6 Routing Protocol for LowPower and Lossy Networks (RPL) \cite{RPL} for communication between constraint-oriented devices. RPL provides support for point-to-point and multipoint-to-point and point-to-multipoint traffic patterns. 

The Light-Weight Implementation Guidance (LWIG) \cite{lwig} working group is focusing to build minimal and
interoperable IP protocol stack for constraint-oriented IoT devices. 
The Thing-2-Thing Research Group (T2TRG) \cite{t2trg} aims to explore the factors which will influence the process
of turning IoT into reality. T2TRG will investigate and list the issues to form the Internet through which low power
constraint-oriented devices can communicate with each other using M2M communication style and with the global Internet. 

\begin{table}[]
\centering
\caption{IoTs Phases and Corresponding Technologies}
\label{table-IoT-LifeCycle-References}
\begin{tabular}{|c|l|l|}
\hline
\multicolumn{2}{|c|}{IoT Phase}                                            & \multicolumn{1}{c|}{Components and Reference(s)} \\ \hline
\multicolumn{2}{|c|}{\multirow{5}{*}{Acquisition and Sensing}}               & RFID\cite{chen2013rfid}                                          \\
\multicolumn{2}{|c|}{}                                                     & WSN \cite{PereraZaslavskyChristenEtAl2014, RazzaqueMilojevic-JevricPaladeEtAl2016}                                            \\
\multicolumn{2}{|c|}{}                                                     & Bluetooth\cite{CollottaPau2015}                                      \\
\multicolumn{2}{|c|}{}                                                     & NFC\cite{WantSchilitJenson2015}                                            \\
\multicolumn{2}{|c|}{}                                                     & UWB \cite{MroueHeddebautElbahharEtAl2012}                                            \\ \hline
\multirow{11}{*}{Data Transmission}         & Current                      & Ethernet\cite{ChristensenReviriegoNordmanEtAl2010}                                       \\
                                            & Enabling                     & Wi-Fi\cite{19ieee802ah,21}                                           \\
                                            & Technologies                 & Wi-MAX                                         \\
                                            &                              & MANETs\cite{RamrekhaAdigunLadasEtAl2015}                                         \\
                                            &                              & Cellular Networks\cite{JoverMurynets2015, JermynJoverMurynetsEtAl2015, NovoBeijarOcakEtAl2015}                              \\
                                            &                              & Satellite Networks \cite{SanctisCiancaAranitiEtAl2016}                             \\ \cline{2-3} 
                                            & Future Enabling              & CRN\cite{AijazAghvami2015}                                            \\
                                            & (or Enabled by IoTs)         & VANETs \cite{HuoTuShengEtAl2015}                                         \\
                                            & Technologies                 & 5G \cite{SkoubyLynggaard2014}                                             \\
                                            &                              & ON\cite{BoldriniLeeOenenEtAl2014}                                             \\
                                            &                              & PLC\cite{OliveiraReisRodriguesEtAl2015}                                             \\ \hline
\multicolumn{2}{|c|}{\multirow{2}{*}{Data Processing and Info. Management}} & Cloud Computing\cite{BottaDonatoPersicoEtAl2014}                                 \\
\multicolumn{2}{|c|}{}                                                     & Big Data\cite{54}                                       \\ \hline
\multicolumn{2}{|c|}{\multirow{3}{*}{Action  and Utilization}}              & Semantics\cite{Al-FuqahaGuizaniMohammadiEtAl2015,KotisKatasonov2013}                                       \\
\multicolumn{2}{|c|}{}                                                     & Actuators\cite{Al-FuqahaGuizaniMohammadiEtAl2015}                                      \\
\multicolumn{2}{|c|}{}                                                     & Applications\cite{iot2, Al-FuqahaGuizaniMohammadiEtAl2015}                                   \\ \hline
\end{tabular}
\end{table}

Moreover, the European Telecommunications Standards Institute (ETSI) \cite{etsi-iot}] is working on the standardization of data security, management, processing and transport for IoT based on IPv6. However, more details about IoT projects and protocols can be found in \cite{iot-IP-protocols}. Nonetheless, the above-mentioned
projects for IoT architecture lies under `all-IP architectures' umbrella.
        
       Furthermore, IP-based networking is inherently designed for host-to-host communication where location (e.g., address) of host plays a vital role, but this location-dependent design creates certain bottlenecks such as efficient information retrieval and delivery. Also, IP networking requires additional protocols to support privacy and security of sensitive data, scalability, mobility, and heterogeneity of nodes. Consequently, traditional IP-based networking is less suitable for these IoT devices and applications. Hence, to provide efficient connectivity among low power IoT devices, a novel networking model like ICN holds much potential \cite{AmadeoCampoloIeraEtAl2014}. Due to this, IETF has also started ICN research group that will help to evolve IP-based architecture \cite{icn-rg}. 
       
              %
       \begin{figure}[!t]
\centering
\includegraphics[scale=0.25]{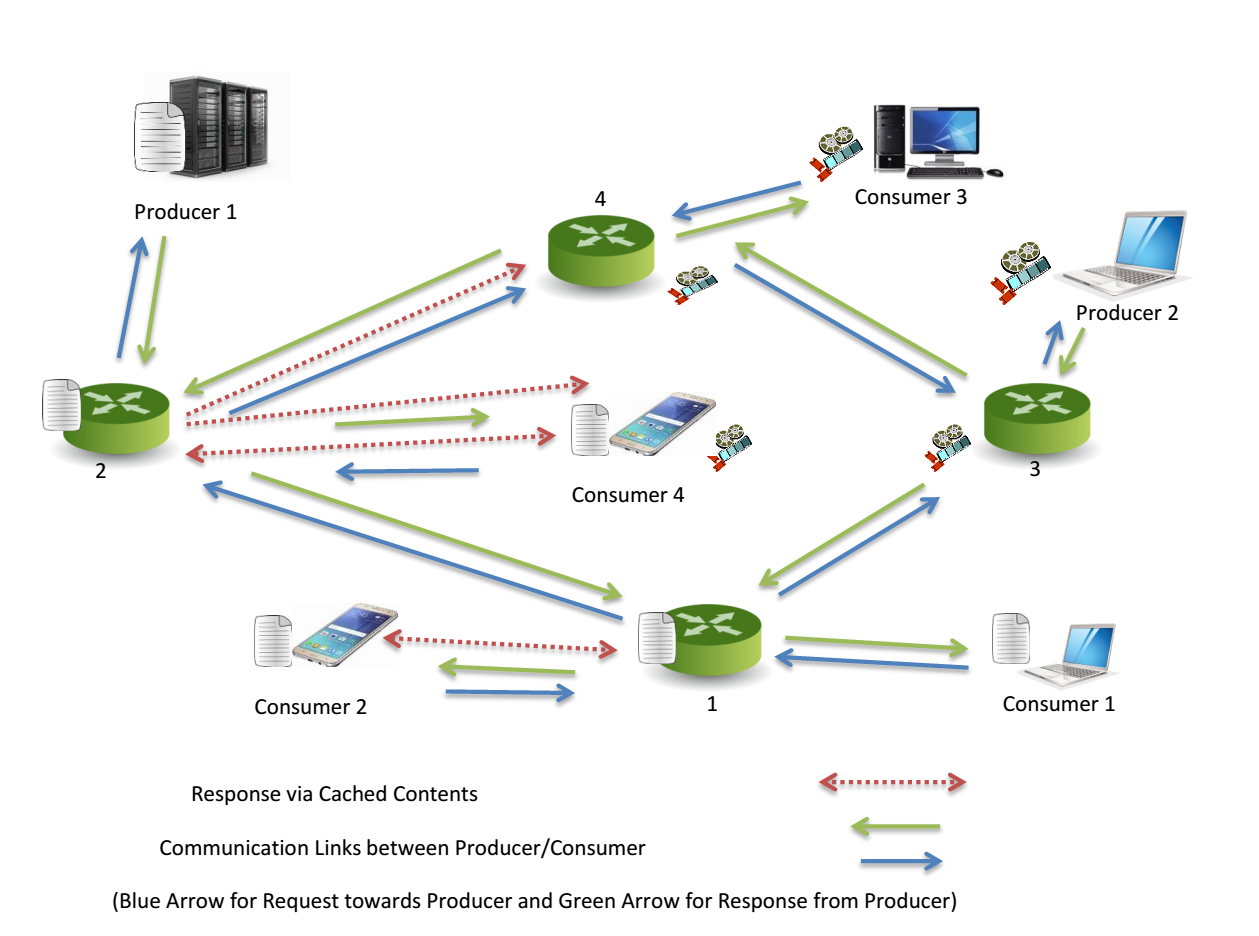}
\caption{ICN Operation: Consumer Requests for a Specific Content by Nearest Routers (1,2,3,4) and Producer Replies and Intermediate Nodes Caches that Content and Fulfils Further Request through Cached Contents Rather Than Sending Request Towards the Original Producer}  
\label{icn1}
\end{figure}
\subsection{Limitations of TCP/IP Architecture and Importance of ICN for IoTs}
From both, today's Internet and IoT context, as all users need data even without knowing the producer of that data. More specifically, in IoTs (i.e., where any specific node can act as producer and consumer at the same time), for example; when an accident occurs somewhere on any road, that vehicle want to inform incoming vehicles about this incident. As a result, flash crowd occurs because only one vehicle is providing the data about that incident. Besides, flash crowds are also the apparent consequence of today's Internet usage \cite{ICNsurvey2012,ICNsurvey,GhodsiShenkerKoponenEtAl2011,GabrielMBrito2013,DannewitzDAmbrosioVercellone2013}.
Flash crowd is a situation which occurs on the Internet when a large number of Internet users request for a particular information item. As a consequence, flash crowds increase network traffic for any particular server (i.e., originating and providing that specific information item) \cite{ari2003managing}. Moreover, data can become unavailable due to end of battery life of many sensors located in that producer vehicle. To minimize flash crowd, ICN provides and supports a much-needed characteristic named: \textit{in-network caching} which minimizes traffic load on the original data producing server while caching the data on intermediate routers. With the help of ICN in-network caching, intermediate routers (any vehicle) can provide data on behalf of the original producer who cached that information item while reducing so-called flash crowd situation. ICN offers in-network caching which makes it more ideal for low power devices. 
\begin{figure}[!t]
\centering
\includegraphics[width=10cm]{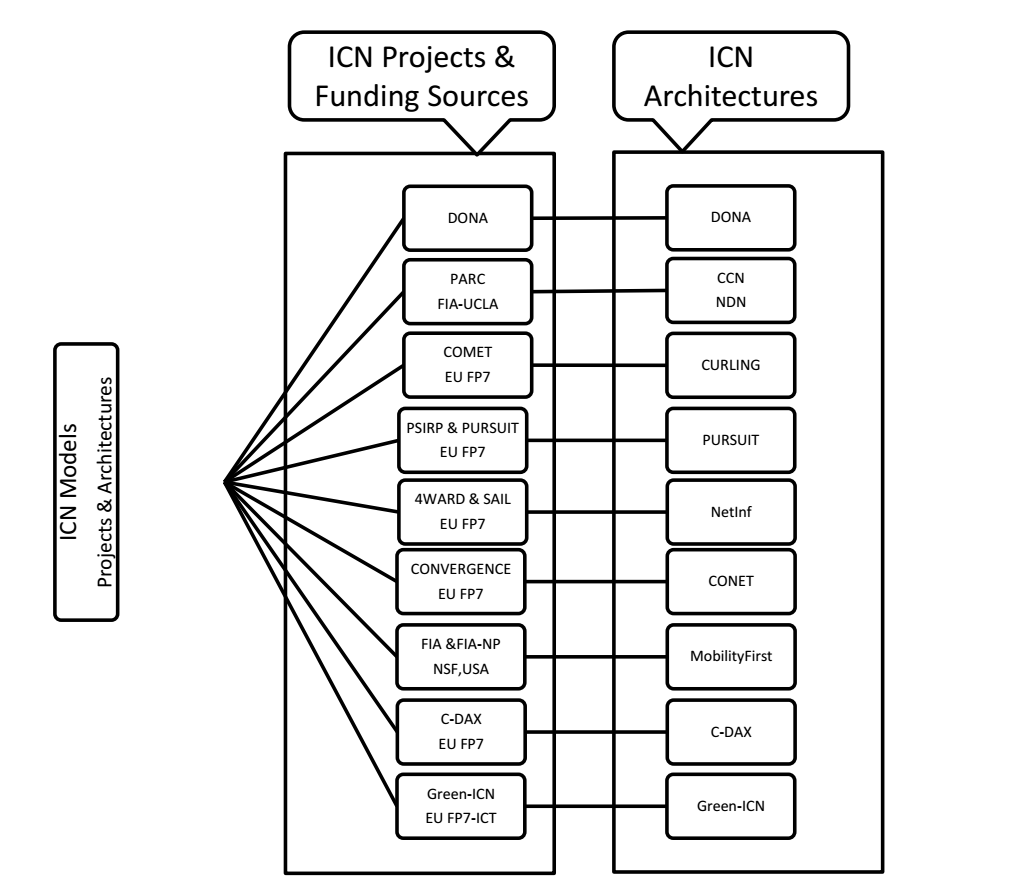}
\caption{ICN Projects, Funding Sources and Architectures}  
\label{ICNProjectsandArchitectures}
\end{figure} 

Moreover, in native ICN, information (i.e., content) is named independent from its location so that it can be located anywhere globally. \textit{Naming} the data and devices make ICN more suitable for IoT as it can combine billions of devices and huge information contents. As ICN supports \textit{receiver-driven communication} making the communication under full control of the receiver. Therefore IoT receiver of information which is more interested in data rather than its location can benefit from it. In addition, push type communication can be provided using beacon messages \cite{majeed2016enabling}. Furthermore, data can only be accessed whenever a receiver explicitly requests a data. As location-independent name searches data, so this provides \textit{opaque communication} between sender and receiver making it more secure. Details of ICN (specifically NDN) operation is shown in Fig.~\ref{icn1}. 

\subsection{IoT Requirements Mapping to ICN Characteristics}
IoT applications which need scalability regarding support for billions of IoT devices and the massive quantity of contents can be build using ICN characteristics like naming the contents, in-network caching and content-based security. ICN naming and name resolution can be efficiently used to provide billions of addresses and names to IoT devices and contents respectively. 
\begin{table*}[]
\renewcommand{\arraystretch}{1.2}
\caption{IoT Requirements Mapping to Supporting ICN Features}
\label{table-Requirement-Mapping}
\centering
\begin{tabular}{c|c|c}
\hline
\bfseries Sr\# & \bfseries IoT Requirement(s) & \bfseries ICN Supporting Features\\
\hline\hline
1. & Scalability	& Naming, In-Network Caching, Content-based Security\\
\hline
2. & Naming and Addressing &	Naming and Name Resolution (Coupled and Decoupled mode)\\
\hline
3. & Mobility &	Decoupled Mode, Naming, Receiver Driven, Location Independence\\
\hline
4. & Security and Privacy &	Naming, Location Independence, Receiver Driven, Content-based Security\\
\hline

5. & Heterogeneity and Interoperability &	Naming and Name Resolution (Coupled and Decoupled mode), Strategy Layer\\
\hline
6. & Data Availability & In-Network Caching\\
\hline
7. & Energy Efficiency & In-Network Caching, Naming\\
\hline
\end{tabular}
\end{table*}

To support IoT applications which involve mobile devices, ICN receiver-driven communication feature along with flexible naming the contents and location independence can play an important role to make hand-off easy for mobile devices. Moreover, ICN in decoupled mode can perform easy re-registration after a hand-off of a mobile device with the nearest new router. 

Security and privacy in IoTs can be provided through following features of ICN. For example, ICN named contents make it easy to inspect that data is flowing according to query, content location independence hides the source of content, receiver-driven communication style confirms that content arrives because the receiver has requested for this content and self-certified contents ensure that the contents are same as sent by the source. 

Heterogeneity among IoT devices can be easily handled when devices named through ICN naming. Different types of IoT devices can operate with each other more efficiently when ICN strategy layer induced in IoT devices. 

ICN in-network caching can enable IoT networks to cache fetched data in (all intermediate) node(s) to enhance data availability in IoT network. Moreover, in-network caching decreases the frequency of fetching data from producer and thus saving network life and making it more energy efficient. Table~\ref{table-Requirement-Mapping} summarizes the mapping of IoT requirements to supporting ICN features. 
\begin{table*}[]
\centering
\caption{ICN Projects, Corresponding Architectures and their Feasibility for IoT}
\label{table-IoT-Feasibility}

\begin{tabularx}{\textwidth}{|m{2cm}|m{1.8cm}|m{8.2cm}|m{4.4cm}|}
\toprule
\textbf{\begin{tabular}[c]{@{}c@{}}Project Name,\\Duration and\\Funding Source \end{tabular}}             & \textbf{\begin{tabular}[c]{@{}c@{}}ICN\\ Architecture\\Name\end{tabular}} & \textbf{\begin{tabular}[c]{@{}c@{}}1. Naming, 2. Caching, 3. Security and 4. Mobility \end{tabular}}                                                                                                                                                                                                                      & \textbf{\begin{tabular}[c]{@{}l@{}}Extent of Suitability for\\  IoT Applications\end{tabular}}                                                       \\ \midrule
DONA 2007 UC Berkeley                                           & DONA                                                                      & 1. Uses flat self-certifying names, which cannot provide scalability. 2. DONA offers both on-path and off-path caching. 3. Self-certifying flat names 4. Early-binding approach    & Not suitable as flat names cannot manage IoT billions of devices data contents                                                                                                            \\ \hline
CCN (2010-2013) by PARC, NDN by NSF and UCLA                        & NDN                                                                       & 1. Provide hierarchical, static and dynamic named data through easy administration. 2. NDN offers both on-path and off-path caching (cache everything) 3. Publisher signature with PKI 4. Listen First Broadcast Later (LFBL) 
                                                                                                 & Highly suitable as IoT devices are constraint oriented, and needs scalable naming technique   \\ \hline
COMET (2010-2012) EU Framework 7 Programme                     & CURLING                                                                   & Unspecified naming scheme, enhance easy access and fast data dissemination through content aware networks, especially supports flash crowds. 2. Works on both on-path and off path through prob-caching). 3. Public key cryptography  4. Specialized mobility-aware Content-aware Routers (CaRs)  & Not suitable \sob{for IoT }as naming scheme is not defined \sob{but suitable for data dissemination applications}\\ \hline
PSIRP and PURSUIT (Sep 2010-Feb 2013) EU Framework 7 Programme & PURSUIT                                                                   & 1. Flat naming provides a decoupled architecture that separates name resolution and data forwarding. 2. Provides effective off-path caching 3. Self-certifying flat names 4. Facilitated by multicast and caching  & Not suitable as flat naming scheme cannot manage billions of IoT devices and data contents \sob{but suitable for data dissemination applications}                                                                                                            \\ \hline
4WARD (2008-2010) and SAIL (2010-2013) EU Framework 7 Programme & NetInf                                                                    & 
1. Flat \sob{self certifying or hashed} naming divides the whole operation in two-steps: name resolution by NRS and data routing by node itself. 2. It offers both on-path and off-path caching 3. Self-certifying flat names with possible explicit aggregation 4. Late Name Binding (LNB) & Not suitable as flat naming scheme cannot manage billions of IoT devices and data contents \sob{but suitable for data dissemination applications}                                                                                                                                                                                                                       \\ \hline
CONVERGENCE (2010-2013) EU Framework 7 Programme & CONET                                                                     & 1. Both (hierarchical and flat Naming) schemes, converges to NDN and DONA in some aspects, designed for multimedia contents, partially dependent on IP-based architecture and partially on ICN-based, 2. Both on-path and off-path caching is provided 3. Publisher signature with PKI 4. Same as NDN with the difference at forwarding information at Border Nodes (BNs) & Not suitable as IoT application requires more than the management of only multimedia contents. IoTs architecture also needs to manage simple contents. \sob{But it is suitable for data dissemination applications}\\ \hline
MobilityFirst FIA (2010-2014) and FIA-NP (2014-to date) NSF, USA & MobilityFirst MF & 1. MF uses flat, self-certifying naming scheme, 160-bit long names to avoid collision and make comparison easy and fast. MF provides best mobility services and employs IP-based architecture in an efficient way 2. MF offers on-path caching 3. Self-certifying flat names  4. Consumer mobility handled using Global Name Resolution Service (GNRS) and Border Gateway Protocol (BGP) for inter-domain routing & Highly required by IoT as it can have both mobile and static devices.
                     \\ \hline

C-DAX FP7-ICT (2012-2016)& C-DAX                & 1. Information is managed in the form of topics using flat and attributes-based naming & For cyber-secure smart-grids and electric vehicles                      \\ \hline
Green ICN (2013-2016) EU Framework 7 Programme & Green ICN G-ICN & 1. Contents can be named by using both flat self-certifying and hierarchical naming schemes with attributes and arranged in topics 2. User assisted caching is employed &Highly required by IoT disaster management and multimedia contents dissemination applications                      \\ \hline

\end{tabularx}
\end{table*} 
\subsection{ \sobia{Feasibility of ICN Models and Projects for IoTs}}
This sub-section presents naming, caching, security and mobility support of nine famous ICN architectures such as DONA \cite{KoponenChawlaChunEtAl2007}, NDN, COMET, PURSUIT, SAIL, CONVERGENCE, MobilityFirst, C-DAX \cite{powergridICN} and Green ICN \cite{greenICN}. ICN major projects and architectures along with funding sources are presented in Fig.~\ref{ICNProjectsandArchitectures} and their feasibility with respect to naming, caching, security and mobility support is summarized in Table~\ref{table-IoT-Feasibility}. However, further details of these architectures can be found in \cite{ICNsurvey}. 

\subsection{ICN-based IoT Architectures}
     \begin{figure*}[!t]
\centering
\includegraphics[width=16cm]{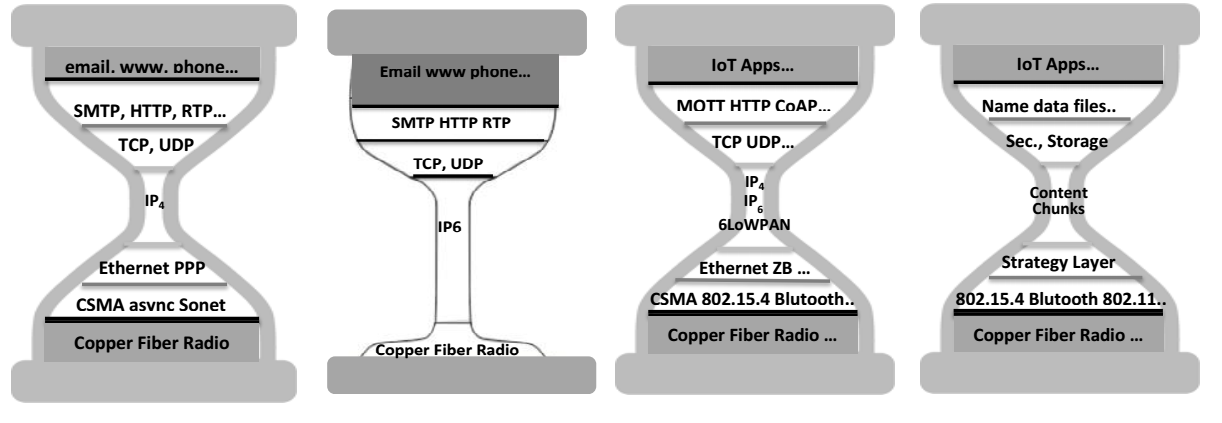}
\caption{IP-based Network Architectures and ICN-based IoT Network Architecture}
\label{icn-iot-arch}
\end{figure*}
   In this sub-section, we present ICN-based IoT research efforts (in following paragraphs) which proposed ICN-IoT network architecture to support IoT needs. 
   The purpose of mentioning these efforts here is not to compare these in any perspective but to showcase the efficient applicability of ICN for IoT along with fertility of this research era.
   
   To build IoT on the basis of ICN, the research community is trying hard. In this context, NDN-based high level \textbf{node architecture} is proposed in \cite{AmadeoCampoloIeraEtAl2014}  to support clean-slate architecture of ICN for IoTs. 
 Three layers NDN-IoT architecture which consists of an application layer, NDN layer and thing layer is presented. Node architecture includes content chunks instead of IP address enabling name-based networking. Strategy layer is introduced to provide transport and forwarding tasks according to access technologies and application needs. NDN operates at the network layer and performs its duty with the help of two planes namely control and management plane and data plane. Control and management plane perform the task like routing, configuration and service models while data plane handles interest and data messages and related jobs like caching strategy. In Fig.~\ref{icn-iot-arch} we present the evolution of Internet architectures. It shows the IP-based architecture, a dedicated version for IoT on the basis of IPv6, extended version (to support IPv4, IPv6, and 6LowPANs) and ICN (NDN) based architecture.
  
To support IoT \textbf{push operations}, three different strategy schemes are presented to provide push-type communication for NDN in \cite{AmadeoCampoloMolinaro2014}. Natively NDN supports pull-based communication, so to provide NDN-based IoT, they provided push support in NDN. First scheme \textit{Interest notification}, modifies interest message by including small data which is to be transmitted. This small data is not meant to be cached. Second scheme \textit{Unsolicited data} transmits a small packet of uData that is not feasible for routing. In third scheme \textit{virtual interest polling (VIP)}, the receiver transmits long live Interests such that whenever data is available, producer replies and on the failure consumer can re-transmit Interest message. They presented the analytical model for Interest notification, Unsolicited data, and VIP and implemented the model in MatLab. VIP outperformed in terms of used network resources and is suitable for massive IoT environment while the other two techniques are suitable for situations where the battery is a critical source.
 
 Furthermore, to provide IoT \textbf{scalability}, CCN (NDN) is identified as the best candidate for IoT rather than RPL/UDP (in IPv6-based 6LowPANs) and implemented in RIOT OS through simulations \cite{BaccelliMehlisHahmEtAl2014}. Wild deployment of ICN is implemented through 60 nodes located in several rooms of several buildings. CCN lightweight version, CCN-lite is simulated and they enhanced CCN through two proposed routing flavors (vanilla interest flooding (VIF) and reactive optimistic name-based routing (RONR)). Both VIF and RONR are evaluated to show that these protocols reduce routing overhead for constraint oriented devices. They also addressed the positive impact of caching and naming the data. 

 Moreover, NDN-based \textbf{secured architecture} (in Python language and Javascripting-based browser to visualize the data) is explored to secure a building. It is installed in UCLA (University of California at Los Angeles)\cite{ShangDingMarianantoniEtAl2014}. Name-based and encryption-based access control method is proposed and implemented to secure sensitive data. This is an initial prototype to showcase the scalability and security performance achieved by NDN instead of IP-based security systems. 
 
 To address and target IoT \textbf{heterogeneity} in terms of both static and mobile devices, a unified ICN-based IoT platform is discussed in \cite{LiZhangRaychaudhuriEtAl2014}. NDN and MF are selected to cater both static and mobile devices. They provided a comparison between both NDN and MF through building management and bus management system scenarios. Different sensors and actuator are considered as static devices while buses are considered as mobile devices. They argue and found that MF outperforms NDN when mobile objects like buses are involved while NDN outperforms MF only when static devices are involved. They have implemented NDN and MF in NS3.
 
 In the following four sections, we categorize and present ICN-based IoT research through ICN caching, naming, security and mobility support which is explored for IoT environment.

\section{ICN-IoT Caching Schemes}
Inherently, the current Internet is designed to forward all requests of same content towards original producer which increases network load, retrieval delay and bandwidth consumption. Moreover, the current Internet lacks support for data dissemination and fast retrieval of the content. These issues raised the need for in-network caching. To cope with these shortcomings of the current Internet architecture, \textit{Content Delivery Networks} (CDNs) were introduced. By employing CDNs, caching is deployed as an overlay patch at the application layer (web-caching) of the current Internet architecture. CDNs are costly to implement and do not utilize network resources efficiently in case of dynamic flash crowds. Thus, in the design of the future Internet architecture caching is added as an essential feature.

 In ICN-based future Internet architectures, caching is implemented at network layer that directly operates on named information. ICN architectures DONA, NDN, SAIL and MobilityFirst primarily support on-path caching while PURSUIT, COMET and CONVERGENCE support both on-path and off-path caching \cite{ICNsurvey}. 

In ICN-based IoT, caching is highly required to disseminate information quickly towards edge devices in a cost-efficient way. As some IoT applications need fresh contents with some specific timing requirements. Moreover, mostly IoT contents are ephemeral in nature that need to replace with the newer versions, for instance, the temperature value of a room needs to be monitored and updated continuously. Furthermore, as IoT nodes are highly heterogeneous, which may differ in the processing resources (i.e., constraint-oriented and powerful nodes) and IoT networks are a mixture of wired and wireless technologies. 

In IoTs, caching at intermediate devices or routers offers many benefits. As the receiver is dissociated from original producer, therefore by caching the contents, security improves and scalability of IoTs network increases \cite{AmadeoCampoloIeraEtAl2014}. Energy efficiency of constraint-oriented devices can be improved and mobility can be handled in more better ways \cite{AmadeoCampoloQuevedoEtAl2016}. Resiliency and life of IoT networks can be improved by employing caching carefully \cite{SourlasTassiulasPsarasEtAl2015b}. 

As caching offer many advantages, it also puts same restrictions and complications on the design of caching strategies for an environment like IoTs. To design ICN-based caching for IoTs, caching strategies must count for some properties of content to cache and node that intends to cache it. Content properties can include popularity, freshness, ephemerality, timing and specific producer while caching node properties can count for battery (power level), the distance of a node from the producer (or/and consumer) and remaining memory. On the basis of these mentioned observations, we provide caching placement strategies into following three categories: 
 
1)    Content-Based Caching (CBC), these strategies decide what content to store on the basis of content properties.

2)    Content and Node-Based Caching (CNBC), these schemes decide whether a node should cache content or not, depending on both content properties and node resources (like battery life). 

3)  Alternative Caching Schemes, algorithms that include the distance of a node from producer or position/role in the network in caching decision lies in this category. ICN-based caching node architecture and cache coherency are also discussed in this category.

An overview of ICN-based caching schemes for IoTs is presented and summarized in Table~\ref{Table-ICN-IoT-Caching}. 
A caching strategy is further divided into following three phases: 

1) Content placement into the cache, in this phase cache space is allocated to contents on the basis of content and/or node. Content placement schemes include like cache each and everything (universal caching), and probabilistic caching. 

2) Content replacement from the cache, in this second phase, when the cache becomes full with contents and there is no space vacant for next upcoming content, it is decided to which already existing content it will replace. Content replacement schemes include like LRU (Least Recently Used) and LFU (Least Frequently Used). 

3) Cache coherency of contents in the cache, in this phase, the validity of contents which reside in the cache, is checked.

Caching performance measures include retrieval delay, hit ratio, network lifetime (how long network will exist in terms of connectivity), interest re-transmissions (total number of interest sent to get a content) and energy consumption per content (how much energy is required to decide about cache a content and/or replace it). 
\begin{figure*}[]
\centering
\includegraphics[width=19cm]{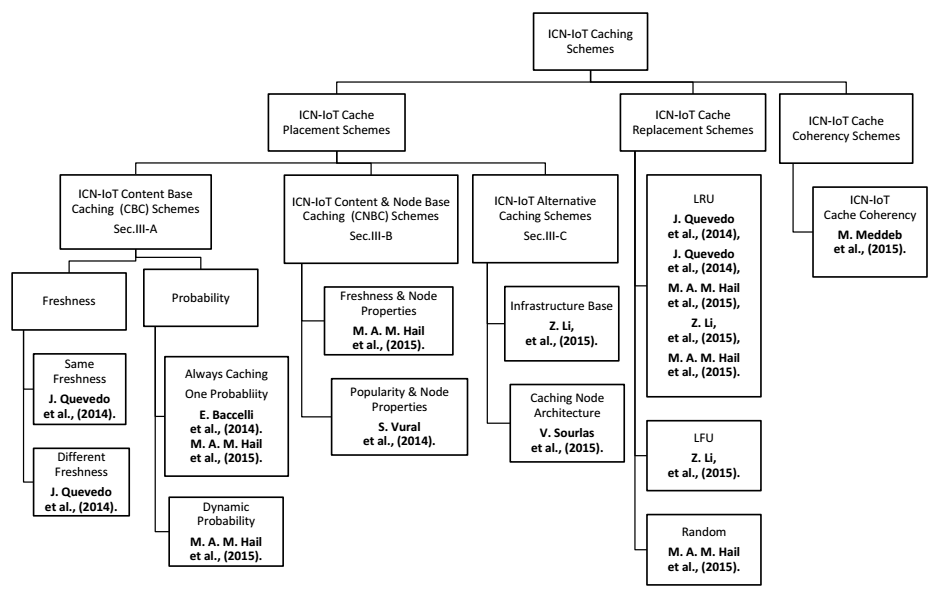}
\caption{ICN-IoT In-Network Caching is Illustrated in Three Phases: Caching Placement, Replacement and Coherency Schemes. Caching placement Schemes are Further Arranged into Three Categories: Content-Based Caching (CBC), Content and Node-Based Caching (CNBC) and Alternative Caching Approaches}  
\label{ICN-IoT-Caching}
\end{figure*}
 ICN-based caching placement methods have been extensively investigated in the context of IoTs in \cite{QuevedoCorujoAguiar2014a}, \cite{QuevedoCorujoAguiar2014}, \cite{HailAmadeoMolinaroEtAl2015}, \cite{LiPointCiftciEtAl2015}, \cite{HailAmadeoMolinaroEtAl2015a} as depicted in Fig.~\ref{ICN-IoT-Caching}. In the following subsections, we survey caching placement schemes along with caching replacement schemes. According to Fig.~\ref{ICN-IoT-Caching}, we sub-classify caching placement schemes into three categories: Content-Based Caching (CBC), Content and Node-Based Caching (CNBC) and alternative caching schemes.

We further classify CBC on the basis of freshness, probability and CNBC schemes according to freshness, popularity along with node properties. We sub-classify alternative caching schemes into infrastructure-based caching, caching node architecture and cache coherency. 

\begin{table*}[]
\centering
\caption{Caching Schemes for ICN-based IoTs according to the classification presented in Fig.~\ref{ICN-IoT-Caching}. CBC is for Content-Based Caching and CNBC is for Content and Node-Based Caching.}
\label{Table-ICN-IoT-Caching}
\begin{tabular}{|c|c|c|c|l|l|c|}
\hline
\toprule
\multicolumn{7}{|c|}{CBC Placement Schemes for ICN-IoT}   \\ \hline 
\toprule
\begin{tabular}[c]{@{}c@{}}Reference\end{tabular} & \begin{tabular}[c]{@{}c@{}}Placement Sub-\\Category Scheme\end{tabular}       & \begin{tabular}[c]{@{}c@{}}Replacement \\ Scheme\end{tabular}                                & Architecture                                                       & \multicolumn{1}{c|}{Comparison}                                                                                                                          & \multicolumn{1}{c|}{\begin{tabular}[c]{@{}c@{}}Parameters\\ Evaluated\end{tabular}}                                                                & Simulator                                                                                 \\ \hline

\cite{QuevedoCorujoAguiar2014a}                                                            & \begin{tabular}[c]{@{}c@{}}Different \\Freshness\end{tabular}                            & LRU                                                                                          & CCN                                                                & \multicolumn{1}{c|}{IP}                                                                                                                                  & \begin{tabular}[c]{@{}l@{}}1.BW Consumption\\ 2.Energy Consumption\end{tabular}                                                                    & \begin{tabular}[c]{@{}c@{}}ndnSIM for CCN\\ and NS3 for IP\end{tabular}                     \\ \hline
\cite{QuevedoCorujoAguiar2014}                                                            & \begin{tabular}[c]{@{}c@{}}Same\\ Freshness\end{tabular}                       & LRU                                                                                          & CCN                                                                & \multicolumn{1}{c|}{-}                                                                                                                                   & \begin{tabular}[c]{@{}l@{}}1.Cache Hit Ratio\\ 2.Avg. number of hops\end{tabular}                                                                  & \begin{tabular}[c]{@{}c@{}}ndnSIM and\\ NS-3 for CCN\end{tabular}                         \\ \hline
\cite{HailAmadeoMolinaroEtAl2015}                                                            & \begin{tabular}[c]{@{}c@{}}Dynamic \\ Probability\end{tabular}                       & \begin{tabular}[c]{@{}c@{}}LRU,\\ Random\end{tabular}                                        & NDN                                                                & \begin{tabular}[c]{@{}l@{}}1.Always Caching\\ 2.Probabilistic Caching\end{tabular}                                                                       & \begin{tabular}[c]{@{}l@{}}1.Hit Ratio\\ 2.Retrieval Delay\\ 3.Interest Re-transmission.\end{tabular}                                              & \begin{tabular}[c]{@{}c@{}}ndnSIM and\\ NS-3 for NDN\end{tabular}                         \\ \hline
\cite{BaccelliMehlisHahmEtAl2014}-\cite{HailAmadeoMolinaroEtAl2015}                                                            & \begin{tabular}[c]{@{}c@{}}Constant Probability \\ (One Probability)\end{tabular} & -                                                                                            & CCN                                                                & \begin{tabular}[c]{@{}l@{}}1.Always caching\\ 2.No caching\end{tabular}                                                                                  & \begin{tabular}[c]{@{}l@{}}Number of packets \\ sent(Interest and Data)\end{tabular}                                                               & RIOT OS                                                                                   \\ \hline
\toprule
\multicolumn{7}{|c|}{CNBC Schemes for ICN-IoT}   \\ \hline
\toprule
\cite{HailAmadeoMolinaroEtAl2015a}                                                           & \begin{tabular}[c]{@{}c@{}}Freshness and\\ Node Properties\end{tabular}          & LRU                                                                                          & NDN                                                                & \begin{tabular}[c]{@{}l@{}}1.No Caching\\ 2.P(.5) Caching \\ 3.Cache each and\\ everything\end{tabular}                                                  & \begin{tabular}[c]{@{}l@{}}1.Hit Ratio \\ 2.Network Life Time\\ 3.Retrieval Delay\end{tabular}                                                    & \begin{tabular}[c]{@{}c@{}}ndnSIM and\\ NS-3 for NDN\end{tabular}                         \\ \hline
\cite{VuralNavaratnamWangEtAl2014}                                                           & \begin{tabular}[c]{@{}c@{}} Popularity and \\ Node Properties\end{tabular}         & Not mentioned                                                                                & Not mentioned                                                      & Not mentioned                                                                                                                                            & \begin{tabular}[c]{@{}l@{}}1.Cost Saving Ratio\\ 2.Hop Distance Ratio\end{tabular}                                                                 & \begin{tabular}[c]{@{}c@{}}MatLAB for\\ Analytical\\ Modeling\end{tabular}                \\ \hline
\toprule
\multicolumn{7}{|c|}{Alternative Caching Schemes for ICN-IoT}   \\ \hline
\toprule
\cite{MeddebDhraiefBelghithEtAl2015}                                                            & \begin{tabular}[c]{@{}c@{}} Infrastructure\\ Based Caching\end{tabular}              & LRU                                                                                          & ICN                                                                & \begin{tabular}[c]{@{}l@{}}1.LCE\\ 2.LCD\\ 3.Prob Caching\\ 4.Betweenness \\ Centrality (Btw)  \\ 5.Client Cache  With \\ Zipf distribution\end{tabular} & \begin{tabular}[c]{@{}l@{}}1.Percentage of validity\\ 2.Response Latency\\ 3.Hop Reduction Ratio\\ 4.Server Hit Reduction Ratio\end{tabular}      & \begin{tabular}[c]{@{}c@{}}Analytical\\ Modeling\\ Simulator Not\\ Mentioned\end{tabular} \\ \hline
\cite{LiPointCiftciEtAl2015}                                                            & \begin{tabular}[c]{@{}c@{}} Infrastructure\\ Based Caching\end{tabular}              & \begin{tabular}[c]{@{}c@{}}LFU in edge\\ routers and LRU\\ in centralized nodes\end{tabular} & \begin{tabular}[c]{@{}c@{}}CCN \\ COMBO\\ project FP7\end{tabular} & \begin{tabular}[c]{@{}l@{}}Current Transparent\\ caching\end{tabular}                                                                                    & \begin{tabular}[c]{@{}l@{}}1.No.of interests sent\\ towards producer Vs\\ towards cache\\ 2.Play-back continuity \\ 3.Average Latency\end{tabular} & OMNET ++                                                                                  \\ \hline
\end{tabular}
\end{table*}
\subsection{Content-Based Caching (CBC) for ICN-IoT}
Most of IoT applications which process the contents put severe constraints on the contents. Some IoT applications demand contents with freshness constraints while other may demand the content with high probability. The probability of content can be set according to the popularity or in a random fashion. In this section, we present ICN-based caching strategies for IoTs which include such content properties in caching decision.  
\subsubsection{Freshness of Content}
IoT contents required by IoT applications are transient in nature which update their values continuously (e.g., temperature sensors update their values and consumer could request the most recent value or with specific date or time). Updated information can be received through specifying freshness value. Thus, caching strategies dealing with freshness are highly important for ICN-based IoTs. In the following subsections, we present those approaches which consider freshness in ICN-based caching design for IoTs.   
\paragraph{Specific freshness Caching}

In \cite{QuevedoCorujoAguiar2014a}, a freshness-based caching scheme is proposed to facilitate consumer applications which inquire for contents with specific freshness values. The consumer has to specify the freshness requirement of the value it needs. Intermediate routers or producer can set (or even can change) the freshness value for the required content raising DoS attack. In Content Store (CS), a new field to set freshness and a check to compare the time stamp of cached data with the requested by consumer have been added to the existing CCN. The consumer is assumed to send a request for the same content and with specific freshness values. Interest packet has been modified by adding a new field freshness parameter. Producer nodes are Wi-Fi nodes connected to Access Points (AP). LRU has been applied as a cache replacement strategy. Freshness value added more control of the consumer in the quality of data being fetched. By adding, a ratio of active time of restrictive in freshness consumer to active time of less restrictive in freshness consumer, caching performed better for IoT applications that need recent data. However in \cite{QuevedoCorujoAguiar2014a} only caching scheme has been presented.  

\paragraph{Caching with same freshness}
IoT environment needs and corresponding ICN features are discussed in \cite{QuevedoCorujoAguiar2014}. Bandwidth and energy consumption are measured for CCN-based IoT scenarios through varying number of nodes (both consumers and producers) and compared against IP. CCN data packets are modified by including both the freshness of content and fraction of size of CS. NS3 based ndnSIM have been used for IP and CCN respectively. Application for the consumer is implemented in the way \sob{that it requests} for the same data from different producers. Total one hundred nodes are included in the simulation \sob{while half of these nodes are producers and half were consumers}. IP-based producers are assumed as WiFi mobile nodes connected to AP, while ICN consumer nodes are set to inquire data of same freshness value. LRU is applied as a cache replacement scheme and cache placement scheme has been designed to include freshness and variable CS size fraction. Impact of increasing sensors requires more bandwidth rather than the increasing number of consumers. This approach is good for IoT scenario where the number of consumers is uncontrollable (e.g., hotspot or flash-crowd). They have found that IP-based case consumed more bandwidth than CCN. Impact of freshness has reduced performance assumed to achieve through caching. To enforce caching, a small CS would be enough if freshness is highly required. However, considered IoT scenario has a fixed number of nodes and implementation is not performed for dynamic IoT scenario. 
\subsubsection{Probability of content}
Some IoT applications which require mix contents from the multiple or single producer(s) like in smart traffic, a car owner may be interested in the traffic condition ahead, the temperature of that area, the exact location of the vehicle and map towards its destination. Therefore, ICN-based caching strategies for IoTs should include factors to cope with these application requirements. In this context, random probability assignment can provide diversity in cached contents.        
\paragraph{Always and Probabilistic Caching}
In \cite{HailAmadeoMolinaroEtAl2015} authors have implemented NDN for IoTs and applied Always and Probabilistic (with P=0.5) caching schemes. LRU and Random replacement algorithms have been applied as cache replacement schemes. Simulations were performed in ndnSIM and NS-3. Total of 36 nodes were included in the simulation, out of which, four were destined as consumers and six were randomly selected as producers in a 400m X 400m area. Probabilistic caching scheme and LRU cache replacement scheme, in a combination, achieved higher results for the cache hit ratio, retrieval delay and interest re-transmissions. Cache size has been varied from 1-4KB but optimal results were achieved when CS size was 4KB. Probabilistic caching and LRU replacement scheme ensured content diversity and most recent contents in the IoT network which are important requirements of IoTs. Though, authors have found caching (even with small CS) beneficial for IoTs.  

In \cite{BaccelliMehlisHahmEtAl2014} impact of Always caching \sob{(Where P is always 1),} is evaluated on RIOT OS \cite{BaccelliHahmGunesEtAl2013} for a large building. They argue through their results that caching is highly beneficial for devices having small memory. Authors support in-network caching for IoTs because it saves bandwidth and energy consumption.   
\subsection{Content and Node-Based Caching (CNBC) for ICN-IoT}
In this sub-section, we survey ICN-based caching schemes that include both content and node \sob{parameters. Content properties like freshness, popularity and node important parameters like battery level, cache size, node location and role in the network are considered for constraint-oriented IoT devices.} 
\paragraph{Probability of Freshness and Node Properties-Based Caching}

  In \cite{HailAmadeoMolinaroEtAl2015a}, the authors presented a probabilistic CAching STrategy for the INternet of thinGs (pCASTING), a caching mechanism considering content property (freshness) and node properties (battery level and cache occupancy). For caching replacement, LRU has been implemented. pCASTING has been compared against cache each and everything (CEE), probabilistic caching (P=0.5) and without caching. Simulations were performed in ndnSIM and NS-3. Total 60 mobile nodes were included in the scenario. There was only one producer and eight consumers were selected. pCASTING achieved a higher cache hit ratio and received data packets by the consumer. Retrieval delays were less than probabilistic and no caching but higher than CEE. However, only one producer has been assumed to \sob{reply.} Popularity of content was not present in the cache decision.
 \paragraph{Popularity and Node Properties-Based Caching}
 
 In \cite{VuralNavaratnamWangEtAl2014}, a caching scheme has been proposed using data freshness, request rate and router properties. Routers have been assigned the task to compute the probability of content, using content properties (freshness and request rate (popularity)) and node properties (incoming request rate and location of the node in the network). Numerical evaluation has been presented in Matlab. However, the proposed caching scheme is for multimedia contents (40GB link has been mentioned in simulation parameters) which requires extensive calculations. Hence it is less suitable for IoT low power, constraint-oriented devices to perform such complex and power-consuming calculations. Moreover, as mobile nodes change locations frequently (network topology changes), the proposed method is highly suitable for static devices. As static devices do not face battery issues to perform such extensive calculations. However, they have not discussed any caching replacement algorithm. 
\subsection{Alternative Caching Schemes for ICN-IoTs}
In this section, we provide a comprehensive overview of caching schemes which do not focus on a particular method (i.e., content or node-based caching) but present caching schemes for IoTs from other perspectives. We categorize these ICN-based caching methods for IoTs into overlay caching and cache coherency schemes because they provide caching network architecture on the existing Internet and cache coherency mechanism for ICN-IoTs. Although ICN-based caching-node-architecture presented in \cite{SourlasTassiulasPsarasEtAl2015b}, is not specifically for IoTs, but we include it to cope with the IoTs disaster management.
\subsubsection{Overlay Caching for ICN-IoTs}
 An overlay shared caching scheme based on ICN is presented in \cite{LiPointCiftciEtAl2015}. Content management (CM) layer is introduced in Fixed and Mobile Converged (FMC) network architecture. This CM layer can be controlled through a network provider or content producer. CM layer decides where content can be cached using its cache and metadata management schemes. Unified Access Gateway (UAG) node stores and forwards the content to any requesting node in FMC network while the network is responsible for transmission of content. A cache controller (CC) is integrated with UAG which provides optimal caching and pre-fetching plans. HTTP traffic passes through this overlay caching. A \textit{Config} packet is added in the CCNx to carry information about caching and cache replacement scheme. Updated CCNx provides transparent overlay caching and in pre-fetching process CC sends \textit{Config} packet to cache node and which in return sends \textit{Interest} message to overlay cache and overlay cache respond with \textit{Data} packet. To provide mobility, they used BonnMotion \cite{bonnmotion}. Better performance of a system is achieved in terms of less number of packets sent towards original server as more packets get a response from overlay caching, average latency and uninterrupted playback than the current system. Presented caching strategy and management scheme offers Caching as a Service (CaaS).
  \subsubsection{Client-Cache and Cache Coherency for ICN-IoTs}
The work in \cite{MeddebDhraiefBelghithEtAl2015} presents an ICN-based cache coherence algorithm and a client-based caching strategy for M2M. Client-cache is named to represent the fact that content is saved in node near to the client node. Authors proposed a client-based on-path caching strategy with less number of nodes and by using nodes that were close to the receiver. 
A cache coherence algorithm has been presented to check the validity of contents. Proposed cache coherence method used expiration-based coherence with variable time expiration for every content. Client-based caching strategy was compared against Leave Copy Everywhere (LCE), Leave Copy Down (LCD), Probability caching and Betweenness Centrality. Client caching along with coherence algorithm has achieved better results in terms of hop reduction ratio, server hit reduction ratio, response latency and validity percentage of contents. To the best of our knowledge, this is only one paper that investigates cache coherency for ICN-based IoTs. However, cache size that is selected, is much larger to suit for low memory devices to hold a large number of contents. Moreover, the discussion about IoT applications which require fresh contents is missing in the proposed method.  
 \subsubsection{Caching Node Architecture for Disaster Management}
Authors in \cite{SourlasTassiulasPsarasEtAl2015b} considered the disaster situation and presented the solution to recover data through cache enabled nodes. A caching scheme is presented to collect fragmented data when a network has fragmented, or some device (producer) has left the current network. They have modified traditional CCN by introducing Satisfied Interest Table (SIT). An expression is presented to show until when content can be available and calculate its disappearance time. It is specifically designed when the producer is moved and network got fragmented (disruptive Scenario). They tried to prolong a content availability through in-network caching because connectivity between friends and family is more crucial and as a result bulk of data is produced in such situations.
NDN router architecture is modified by augmentation of SIT. SIT keeps track of users with the same interests and get the required data. SIT is meant to forwards interest packet to users on the basis of entries it has saved. SIT entries are erased only when that user left the network. Interest packet is modified to be satisfied by the producer or satisfied consumer by introducing Distention Flag (DF). If DF is 1, SIT provides the satisfied user with the same interest and now provides the data against requested interest. Data Packet is same as of NDN. However, this scheme requires a lot of memory, so it is natively not suitable for IoT small devices. But intrinsically suitable for nodes with excessive memory and it can be employed somewhere in IoT networks (e.g., as a backup node in IoT disaster management applications). \sob{Moreover, it requires other users willingness to disseminate data and respond to queries that can put a lot of burden on the network management and can raise security issues.}
\subsection{Summary and Insights}
We have surveyed ICN-based caching schemes in the context of IoTs and provided a classification in Fig.~\ref{ICN-IoT-Caching}. We have broadly categorized ICN-IOT caching \sob{mechanism} into \sob{three phases}: caching placement, replacement and coherency phases. Caching schemes have further categorized into three strategies: CBC, CNBC and alternative caching. 

CBC schemes compute properties for every content, \sob{which include freshness and popularity of content. Researchers have put more focus on exploring the content freshness while popularity has been explored in a few approaches.} Therefore, ICN-based content popularity caching for IoTs seeks urgent attention from the research community.
 
On the other hand, it is important to consider both node and content properties while making cache decision. On this side, a few efforts have been made to combine both features in cache placement strategies \sob{\cite{HailAmadeoMolinaroEtAl2015a}-\cite{VuralNavaratnamWangEtAl2014}}. For this caching mechanism, we categorize it into CNBC strategies. CNBC strategies include content properties along with IoT node characteristics like battery timings, CS size, node position and caching module designing in the node and IoT network type. As IoT nodes assumed to have the low processing power, memory, and battery. However, current literature on caching is missing IoTs low power and low memory characteristics of nodes and IoT applications with mobile devices. Moreover, caching strategies lacks in push traffic type consideration for IoT network. 

\sob{In comparison to decide about optimal caching schemes in ICN-based IoTs, CNBC is better than CBC alone regarding throughput, but apparently it requires more resources to compute about caching decision. ICN-based energy efficient caching schemes for IoTs are also needed to explore by the research community.}

Besides both CBC and CNBC, we categorize remaining ICN-based caching schemes for IoTs into alternative caching schemes. This include application specific caching node architecture like disaster management application, cache coherency protocol and overlay caching. This third category is decided irrespective of both node and content properties.  

\sob{The survey proves that CBC has been explored to some more extent than CNBC. This is because CBC protocols directly deal with content properties like freshness and popularity. As every IoT application demands contents with different properties, for example, real-time applications demand highly fresh contents while flash crowds need more popular contents. As a result, CBC schemes are easy to explore for IoTs application scenarios. On the other hand, CNBC schemes are somewhat challenging to implement as ICN-based IoT node and network architecture are still under research and construction phase.} 

In caching replacement strategies, mostly LRU has been implemented in common nodes due to its better results while LFU has been considered for edge nodes. Furthermore, random replacement scheme is easy and simple to implement that ensures high data diversity as well.

So far, there is only one cache coherency protocol for ICN-based IoTs \sob{\cite{MeddebDhraiefBelghithEtAl2015}}. Thus ICN-based coherency protocols for IoTs are highly required to provide content validation in IoT applications.    

In a nutshell, our extensive survey of ICN-IoT caching schemes indicates that ICN caching provides better IoT network performance and improves data delivery. \sob{Future research needs to explore CNBC caching schemes for IoTs constraint-oriented nodes while accommodating both transient and ephemeral contents.}    
 
\section{ICN-IoT Naming Schemes}
Fundamentally the IP-based Internet was designed to communicate between academic devices, but with time Internet usage has expanded from academic communication to fulfil society communication needs. Later on, with the help of add-on and specific purpose patches, IP-based Internet tried to fulfil current needs of society. As a consequence, by adding patches, IP-based Internet architecture provides current needs at the cost of more complex, slow, extra expensive communication and sharing of content. With the time and keeping current expectations from the Internet in mind, researchers proposed the idea of ICN that is based on name-based networking. The named content can be accessed independently irrespective of its location of existence. In ICN, the name of content requested is required instead of sender and receiver address pair. Therefore, this makes ICN as receiver-driven communication model in which receiver is responsible and have full control over whole communication instead of a sender. The network is responsible for and will have to look for content providing best source \cite{ICNsurvey}-\cite{ICNsurvey2012}.       

As users are more and more interested to receive content rather than the location of the content from where it is coming, so ICN approaches provide the ways to name data according to some constraints. The user can get desired contents by only providing their names.

ICN naming can also outperform in naming IoTs contents. IoTs contents are transient in nature and it is undoubtedly possible for one content to have many versions based on time and sensors which generate the same information. 
Moreover, IoTs contents are huge in number like billions of contents are likely expected to produce in any single second and IP-based Internet cannot address 50 Billion \cite{50BDevicesCisco} connected devices efficiently. According to CISCO report, there will be 12.2 Billion IoTs smart and constraint-oriented connected devices in 2020 \cite{M2Mconnections2016-12B}. Also, IoT network architecture is assumed to support scalability and heterogeneity. 

\begin{figure*}[!t]
\centering
\includegraphics[scale=.7]{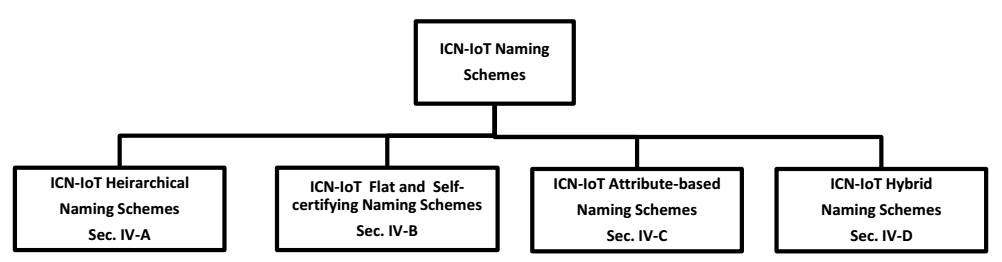}
\caption{ICN-IoT Naming is Categorized into Four Categories: ICN-IoT Hierarchical Naming Schemes, ICN-IoT Flat and Self-Certifying Naming Schemes, ICN-IoT Attribute-based Naming Schemes and ICN-IoT Hybrid Naming Schemes}  
\label{ICN-IoT-Naming}
\end{figure*}   
Mainly there are two naming techniques (hierarchical naming structure and flat self-certifying/hash naming) \sob{which} are available through ICN architectures. CCN \cite{ccn2009} / NDN \cite{ndn2014arcticle} name contents in hierarchical manner while other ICN approaches (DONA \cite{KoponenChawlaChunEtAl2007}, PURSUIT \cite{psirp}, COMET \cite{comet-curling}, MobilityFirst \cite{mobilityfirst}, SAIL \cite{sail} and CONVERGENCE \cite{convergence} ) follow flat self-certifying names. Third naming scheme, attribute-based  has been used initially in CBCB (Combined Broadcast and Content-Based) routing \cite{carzaniga2004routing} and can be used in combination with prior two naming techniques \cite{apr-zhang2016uniform}-\cite{oct-zhang2016uniform}. However, most of the research efforts considered and explored hierarchical naming technique for IoTs \cite{burke2013securing}-\cite{burke2014secure}-\cite{AbidySaadallahyLahmadiEtAl2014}-\cite{BaccelliMehlisHahmEtAl2014}-\cite{shang2014securing}-\cite{AmadeoCampoloIeraEtAl2015}-\cite{AmadeoCampoloQuevedoEtAl2016}. Some researchers focus on hybrid naming schemes which incorporate both hierarchical and flat with attribute-based naming \cite{BoukAhmedKim2014}-\cite{BoukAhmedKim2015}. We categorized ICN-IoT naming schemes into four types which can be visualized in Fig.~\ref{ICN-IoT-Naming}. 

Therefore, naming IoT (devices and) contents through ICN ensure efficient addressing and scalability, more security, better mobility and support for heterogeneous devices \cite{AmadeoCampoloMolinaro2016}-\cite{AmadeoCampoloQuevedoEtAl2016}.
 
\subsection{Hierarchical-based ICN-IoT Naming}
These names are human-readable names and offer name aggregation.  Moreover, NDN and CCN approaches use hierarchical naming. It follows the hierarchical structure to name contents like contents are named on web pages through URLs. Hierarchical naming provides good compatibility with the existing Internet applications and supports name aggregation. Through variable length, hierarchical names are highly scalable which fulfils the ultimate requirement of IoT contents and devices that are huge in number. Searching for a specific name through hierarchical naming already has good compatibility with existing web-browsers architectures. Hierarchical names reduces the routing table information through name aggregation\cite{apr-zhang2016uniform}-\cite{oct-zhang2016uniform}.

On the other hand, long and variable length hierarchical names cause degradation in search efficiency and also for low power devices, it could create more performance degradation. 

In \cite{AbidySaadallahyLahmadiEtAl2014}-\cite{CCNxsaadallah2012} hierarchical content naming scheme is used to name the contents. This work is conducted to design, implement, and integrate a CCN communication layer in Contiki, based on named data for wireless sensors and networking embedded systems. A CCN name is a hierarchical name attributed to content and consists of a simple series of components of arbitrary lengths. No limitations are imposed that what sequence of the bytes will be used. The implemented communication layer specifies only the name structure and does not assign any meanings to names. It is up to applications or global naming conventions to set and interpret meanings given to names. Application developers are free to design their own custom naming conventions.
 However, interest is processed in a hierarchical way. Matching is performed on the prefix to provide multiple responses. 
 \sob{They used CCN for every node.} Contiki OS is used with Cooja simulator to simulate physical TelosB \cite{TelosB} nodes. It is the first paper that implemented CCN in Contiki OS. However, only one sink (consumer) node is considered with ten to forty sensors (producer) nodes. \sob{Only s}tatic nodes are considered. Moreover, the provided naming scheme is not easy to compare for a specific data as hierarchical names are long and complex to perform matching. It is suitable for an IoT application having sensors deployed at fixed places (e.g., Building automation and management). 

Similarly, in \cite{AmadeoCampoloIeraEtAl2015} NDN hierarchical naming scheme is modified for smart homes. Authors have provided namespace specific to home-related tasks. Naming scheme is designed to consist of two part: first for ``\textit{configuration and initialization}" for the smart home application and described by prefix ``\textit{/homeID/conf/}" while second part is for the ``\textit{tasks}" that need to be performed by smart home application and indicated through prefix ``\textit{/homeID/task/}". Tasks are further specified by two named-components, type (is selected from ``\textit{/action}" and \textit{/sensing}) and subtype (is chosen from real tasks like ``\textit{/light, /temp, /airCond}") respectively. Name aggregation is suggested to support task aggregation to reduce the number of sent messages and hence to reduce network bandwidth. However, they did not provide any simulations to show how name\sob{s} are carried by interest and data messages. The proposed naming scheme is designed for the home scenario and thus cannot be used for IoT applications which involve mobile devices.

NDN \sob{hierarchical} naming is explored and deployed for lighting automation by UCLA \cite{burke2013securing}. Contents are named according to three parts: \textit{/constant-namespace/command/randomizer$\|$auth-tag}. For instance, in ``\textit{UetTaxila/CPED/VipLab/Light01/ON/13:15:046FHDK}",  here ``\textit{UetTaxila/CPED/VipLab/Light01/}" represents light numbered as ``01", located in Video and Image \sob{Processing Laboratory} (VipLab) in Computer Engineering Department (CPED) of University of Engineering \& Technology, Taxila (UetTaxila), ``\textit{/ON/}" directs to turn this light ``ON" and ``\textit{/13:15:046FHDK}" indicates the time and corresponding computed hash of the name to ensure security of the content.   

Authors in \cite{BaccelliMehlisHahmEtAl2014} have implemented NDN on IoT constraint-oriented devices for building automation. They have demonstrated the use of small names of size \sob{up to} 12 bytes. They find NDN can support maximum name length \sob{up to} 30 bytes. They believe that hierarchical, short and non-human-readable names are highly suitable for IoT smart devices while maintaining name-aggregation.  

Further, in \cite{shang2014securing} authors believe that hierarchical, human-readable and application-specific names simplify both creation and processing tasks. NDN naming scheme is implemented to secure using ICN for UCLA campus. The designed prototype is implemented in Python and embedded in a browser-based interface. Namespace comprised of main root name followed by two sub-category names. For example, ``\textit{/ndn/ucla.edu/bms/building/Strathmore/data/power/$<$time-stamp$>$}" specifies NDN application deployed at UCLA university for university-building-management-system and fetches power data according to the specified time of Strathmore building located in UCLA. Moreover other sub-namespace, ``\textit{/ndn/ucla.edu/bms/user/public/key/$<$key-id$>$}" directs NDN-based BMS application towards the public user (having multiple keys) through a user-specific key.

However, we argue that short but fixed hierarchical names are suitable for IoT contents because it offers high scalability and name aggregation. Therefore, researchers need to look for the solutions to improve look-up efficiency and optimization of routing table size for IoT constraint-oriented devices.. 
\begin{table*}[]
\centering
\caption{ICN-based IoT Naming Schemes are summarized according to the Fig.~\ref{ICN-IoT-Naming}. Here NLAPB is for Name Lookup solution with Adaptive Prefix Bloom Filter.}
\label{ICN-IoT-Naming-Table}
\begin{tabular}{|c|c|c|l|l|l|}

\hline
\toprule
Reference                                                          & Architecture & Comparison                                                                                                                                           & Parameters Evaluated                                                                                                                                                                      & IoT Application                                                                                & \begin{tabular}[c]{@{}l@{}}Simulator (OS,\\ Programming \\ Platform, Language)\end{tabular}                            \\
 \hline 
\toprule
\multicolumn{6}{c}{Hierarchical Naming Schemes for ICN-IoT}                                                                                                                                                                                                                                                                                                                                                                                                                                                                                                                                                                                                    \\
 \hline 
\toprule

\cite{CCNxsaadallah2012}-\cite{AbidySaadallahyLahmadiEtAl2014} & CCNx         & IP                                                                                                                                                   & \begin{tabular}[c]{@{}l@{}}1.Retrieval Delay \\ with and without \\ caching \\ 2.Number of \\ Exchanged Messages\end{tabular}                                                             & \begin{tabular}[c]{@{}l@{}}Temperature\\ Measurement\\ Wireless\\ Sensor Networks\end{tabular} & \begin{tabular}[c]{@{}l@{}}Contiki OS and\\ Cooja Simulator\end{tabular}                                              \\ \hline
\cite{BaccelliMehlisHahmEtAl2014}                                & CCNx         & \begin{tabular}[c]{@{}l@{}}6LoWPAN/RPL/UDP\\  1.Vanilla Interest\\Flooding (VIF) VS.\\ Reactive Optimistic Name\\-based Routing (RONR)\end{tabular} & \begin{tabular}[c]{@{}l@{}}Number of Consumers\\  VS.\\  Number of Messages Sent\\ (With and without Caching)\end{tabular}                                                                & Building Automation                                                                            & RIOT OS                                                                                                                \\ \hline
\cite{AmadeoCampoloIeraEtAl2015}                                 & NDN          & -                                                                                                                                                    & \begin{tabular}[c]{@{}l@{}}1.Number of transmission(s)\\ 2.Number of Exchanged \\ essages Vs Number \\ of producers\end{tabular}                                                          & Smart Home                                                                                     & \begin{tabular}[c]{@{}l@{}}No simulations\\ Not mentioned\end{tabular}                                                 \\ \hline
\cite{burke2013securing}                                         & NDN          & -                                                                                                                                                    & \begin{tabular}[c]{@{}l@{}}No simulations\\ Not mentioned\end{tabular}                                                                                                                    & \begin{tabular}[c]{@{}l@{}}Light Control System\\(Instrumented\\Environment)\end{tabular}                                                & Not mentioned                                                                                                          \\ \hline
\cite{shang2014securing}                                         & NDN          & -                                                                                                                                                    & \begin{tabular}[c]{@{}l@{}}No simulations \\ Not mentioned\end{tabular}                                                                                                                   & Building Management Systems                                                                    & \begin{tabular}[c]{@{}l@{}}Python-based\\ Application\\ Java-Scripting\\ Data Visualization\\ Application\end{tabular} \\
\hline
\toprule
\multicolumn{6}{c}{Flat ( and Self-Certifying) Naming Schemes for ICN-IoT}                                                                                                                                                                                                                                                                                                                                                                                                                                                                                                                                                                                      \\
\hline
\toprule
\cite{dinh2013potential}                                         & CCNx         & IP-based WSN                                                                                                                                         &  \begin{tabular}[c]{@{}l@{}}1.Average energy\\  consumption  \\ 2.Average delay\end{tabular} & WSN                                                                                            & \begin{tabular}[c]{@{}l@{}}Contiki OS and\\ Cooja Simulator\end{tabular}                                               \\ \hline
\cite{hong2015flat}                                              & ICN          & Not provided                                                                                                                                         & Not provided                                                                                                                                                                              & \begin{tabular}[c]{@{}l@{}}Not for low-power\\  IoT devices\end{tabular}                                                                   & \begin{tabular}[c]{@{}l@{}}No Simulations\\ Not mentioned\end{tabular}                                                 \\ \hline
\toprule

\multicolumn{6}{c}{Attribute-based Naming Schemes for ICN-IoT}                                                                                                                                                                                                                                                                                                                                                                                                                                                                                                                                                                                                 \\ \hline
\toprule

\cite{li2014attribute}                                           & ICN          & \begin{tabular}[c]{@{}l@{}}With and without\\  ontology\end{tabular}                                                                                 & \begin{tabular}[c]{@{}l@{}}1. Storage Overhead \\ 2. Transfer Time Consumption\end{tabular}                                                                                               & Smart Hospital                                                                                 & C Language                                                                                                             \\
\\ \hline 
\toprule

\multicolumn{6}{c}{Hybrid Naming Schemes for ICN-IoT}                                                                                                                                                                                                                                                                                                                                                                                                                                                                                                                                                                                                          \\ \hline
\toprule

\cite{BoukAhmedKim2014}                                          & NDN          & No Comparison                                                                                                                                        & -                                                                                                                                                                                         & Vehicular Ad-hoc Networks                                                                      & \begin{tabular}[c]{@{}l@{}}No Simulations\\ Not mentioned\end{tabular}                                                 \\ \hline
\cite{quan2014social}                                            & NDN          & \begin{tabular}[c]{@{}l@{}}No naming \\ Comparison\end{tabular}                                                                                      & \begin{tabular}[c]{@{}l@{}}1.Start-up delay \\ 2.Playback Freezing Ratio\end{tabular}                                                                                                     &  \begin{tabular}[c]{@{}l@{}}Multimedia Contents\\ dissemination in\\  VANETs\end{tabular}                                                    & \begin{tabular}[c]{@{}l@{}}NS3 with\\ ndnSim\end{tabular}                                                              \\ \hline
\cite{BoukAhmedKim2015}                                          & NDN          & \begin{tabular}[c]{@{}l@{}}1.NLAPB \\ 2.Simple Trie\end{tabular}                                                                                     & \begin{tabular}[c]{@{}l@{}}1.Processing Time to\\  add prefixes \\ 2.Processing Time to\\  delete prefixes\\ 3.Processing Time to\\  search prefixes\\ 4.Memory \\ consumption\end{tabular} & Vehicular Ad-hoc Networks
                                                                     & Not mentioned                                                                                                         
\\
\hline 
\cite{sobia-iot2018}                                          & CCN          & \begin{tabular}[c]{@{}l@{}}Hierarchical and flat\\ naming aggregation\end{tabular}                                                                                     & \begin{tabular}[c]{@{}l@{}}1.Interest transmission\\rate 2. Number of \\covered hops and \\exchanged messages\end{tabular} & IoT Smart Campus
                                                                     & \begin{tabular}[c]{@{}l@{}}Contiki OS \\with Cooja Sim \end{tabular}
\\
\hline 
\end{tabular}
\end{table*}  
\subsection{Flat Self-certifying-based ICN-IoT Naming}
ICN native approaches like DONA \cite{KoponenChawlaChunEtAl2007}, MobilityFirst \cite{mobilityfirst} and \textit{NetInf} \cite{netinf} follows the flat, short and self-certifying names. These names can be computed using the hash of the content or any sub-content as part of it and thus can be non-human-readable. Moreover, flat names can be of any fixed length and therefore simple and easy to process in routing because these names take less computing resources and consume less space to cache.

Although there are very few research attempts which explored ICN flat naming alone. We survey and present these flat naming research efforts in the following paragraphs. Moreover, as such these efforts cannot be used for IoTs but in hybrid form.

In \cite{dinh2013potential}, authors presented ICN flat naming scheme for WSNs. The presented naming scheme has two parts: the first is to identify the category and the second is for content. They have investigated CCN naming in Contiki OS and results indicate that the proposed naming scheme outperform IP regarding energy consumption and delay. As this scheme is for WSN, therefore it can play an important part in IoT applications which involve low-power sensors.

In \cite{hong2015flat} authors present routing scheme based on flat naming. Bloom filter is used to provide name aggregation and efficient searching. They have introduced the concept of containers to save the contents. Controllers controlled containers and accessed through access controllers. Flat names play a great role in the routing of named contents because these are short in length which makes it easy to route and less complex in comparison. However, this work has not involved constraints required by low-power constraint-oriented devices. Hence it is not suitable for all IoT applications.
   
In \cite{namingnrouting2012survey} authors survey naming schemes of ICN architecture and argue that self-certifying names provide name-persistence, security-binding and universal uniqueness.  Moreover,  \cite{adhataraocomparison} provide naming schemes comparison and authors argue that flat names are agnostic to the structure of the data, easy to manage and seems more scalable at the network layer. 
Most of the work regarding flat names is conducted for name base routing\cite{sun2015geometric}-\cite{sun2015geometriccom}.

However, on the other hand, flat names does not provide name-aggregation which is needed for IoT contents and devices to ensure scalability. As flat names can increase the routing table entries making it more complex. Therefore, it will increase the delay in processing a query and will need ample space. Moreover, most of the flat names are non-human-readable, thus to respond any query, a third-party translation mechanism will be required. Furthermore, as IoT devices are small in memory and power, so flat names alone are not suitable for IoT contents and devices.    
\subsection{Attribute-based ICN-IoT Naming}
This naming approach extracts attributes of the content. It is also used initially in CBCB \cite{carzaniga2004routing}. As content attributes can include production date and time, content type, content location, content version number and any specific property of the content. Therefore, this naming approach does not ensure global uniqueness of the content because it is possible to find many responses against a single query and it is hard to find unique content in a short time.
However, it supports searching using easy and known keywords for the content. 
To secure contents, a routing scheme is provided in \cite{ion2013toward} using attributes of the content. 
In \cite{li2014attribute}, an attribute-based naming scheme is presented with the help of ontologies to manage contents in distributed environments. Authors claim that proposed attribute-based naming scheme provides better privacy, simple namespace management and reduces computation cost for the user to determine accessibility. A hospital scenario is presented and described. In our observation this attribute-based accompanied ontologies naming scheme can outperform in IoT applications where privacy is highly needed,  for example, smart-health and smart-transport.

In \cite{namingnrouting2012survey}, authors believe and suggest to use keywords of content created by owner \sob{as they} take less time in searching while making the lookup process easy.
   
For IoT applications, attribute-based naming can help in a perspective that IoT applications are extremely different and the user can specify the required content name in keywords. Attributes can be saved as keywords or hash of attributes to provide more security. Efficient advance search is only possible through attributes of the content. However, fetching unique content seems difficult with only attribute-based naming. To fulfil this other naming schemes can be combined in a hybrid fashion.    
\subsection{Hybrid ICN-IoT Naming}
Hybrid ICN-based naming schemes for IoTs refer to naming schemes which combine three naming schemes or any two of them. The purpose of combining the above-mentioned three naming schemes is to utilize their best features for IoT applications. 
Advantages of these hybrid naming schemes are manifold like improved security, better compatibility, enhanced scalability, and easy name management \cite{apr-zhang2016uniform}-\cite{oct-zhang2016uniform}. 

In \cite{BoukAhmedKim2014}, a scalable naming scheme is proposed for mobile nodes like vehicles and their produced mobile contents. The content name consists of three components:

i) A scheme, ``\textit{vhn}" which specifies the vehicular network or vehicular identifier,

ii) A prefix which is purely a hierarchical component and contains information of producer (car) and details about content,  and

iii) The flat part is the hash of the item-name, owner-name or signature of owner.

However, they did not provide any supporting simulations and feasibility for the proposed scheme. Moreover, the proposed naming scheme based names can be very long and suitable for VANETs only. This scheme is complex for IoT constraint-oriented devices as they can hardly forward/store such long names from/in their CS. 

In \cite{quan2014social}, a hybrid naming scheme is proposed and used for multimedia contents in VANETs using ICN. Proposed naming scheme comprised of following three parts: 

i) Prefix ``\textit{hmn}": indicates ``hierarchical multimedia naming" and hierarchical component names which also used for routing and name-aggregation,

ii) The flat part is the hash computed on the complete name or part of it and
 
iii) Attribute part is the attributes of the content.

 These three parts \sob{(prefix,} flat and attribute) are separated by ``:" while both prefix and attribute sub-components are separated through ``/". This work is designed and evaluated for the dissemination of multimedia contents in VANETs.
 
In \cite{BoukAhmedKim2015}, authors investigate hybrid naming scheme proposed in \cite{BoukAhmedKim2014} and presented their corresponding results for VANETS. Authors claimed that the proposed hybrid naming scheme take less space to save more names as compared to NLAPB \cite{quan2014scalable} and simple trie. They have performed simulations, and results indicate that lookup time and memory management improves for VICN. Maximum prefix allowed length counted as 72bytes. Therefore, this hybrid naming scheme is well suited for low power devices and can support IoT devices when underlying technology is IEEE 802.15.4 Zigbee (i.e., Payload size is 127 Bytes).

In \cite{sobia-iot2018}, we propose a hybrid naming scheme for IoT-based Smart Campus (IoTSC). Hybrid naming scheme names the IoT contents while combining hierarchical and flat components. The proposed naming scheme takes the domain name, location, task as hierarchical component and hash of device name as a flat component. The flat component is computed through FNV-1a hash to maintain the integrity of the content. The proposed scheme is evaluated and simulated for both static and mobile Zigbee devices in Contiki OS with cooja simulator. Results show the better performance is achieved in terms of interest satisfaction rate, the number of covered hops and name-aggregation ratio.   

Through ICN-based hybrid naming, many advantages of the above-described schemes (hierarchical, flat and attribute-based) are expected to improve further while minimizing the effects of different constraints in the case of IoTs.
\subsection{Summary and Insights}
In this section, we have surveyed ICN-based naming schemes which are proposed and investigated for IoT applications. We categorized ICN-based naming schemes for IoT into four categories: hierarchical, flat self-certifying, attribute-based and hybrid naming schemes.

Our survey indicates that for IoTs, NDN (CCN) hierarchical naming schemes and hybrid naming schemes gained more attention from the research community as compared to flat and attribute-based naming schemes. We observe that the main reasons behind NDN (CCN) hierarchical naming feasibility for IoTs are, simple and easy name-aggregation and better support for scalability. Moreover, human-readable hierarchically structured names with unlimited length provide faster searching as compared to other schemes and also name-aggregation saves a lot of space while making routing easy. 

On the other hand, ICN-based hybrid naming schemes enhance the benefits of combined naming schemes. A hierarchical component is added with the aim to provide scalable and efficient name aggregation with less number of entries to make routing process simple and easy. While the flat-name component is concatenated to ensure improved security and privacy. Attributes of content are also included to make fuzzy searching possible through attribute keywords.  

Our survey identified that very few research studies have adopted and investigated flat and attribute-based naming separately for IoTs. Although fixed length, non-human-readable flat naming provide better security and privacy through more easy and simple computations. However, this scheme does not provide better scalability, name-management, and aggregation. Moreover, this is the apparent cause behind less motivation to explore flat naming for IoTs. Though, we highly suggest to use flat names to meet IoTs privacy and security requirements as a name component.

Similarly, attribute-based naming schemes alone gained less attraction from ICN-IoT research community. Attribute-based naming can assist better in advance IoT applications (for instance, an IoT application need temperature values extracted from both node one and ten during the time 04:00 AM to 06:00 AM for any specific date from the desired area) which require contents according to specified features. Thus, we recommend that attribute-based naming should be explored for IoTs.  

 However, to conclude, we recommend that hybrid naming schemes will outperform to name IoT contents and devices accompanying hierarchical, flat and attribute-based naming.



\section{\sobia{ICN-IoT Security Schemes}}
In today's Internet and IoT applications, security is a basic need and a central factor from the design perspective. As almost all IoT applications tend to take data from our daily life gadgets, then with the help of third parties that data is processed. The involvement of third parties creates a potential to affect our privacy. Moreover, content security is not inherited in IP-based Internet applications, but security features like content integrity and device authentication are added later as an add-on. IP-based protocols like EAP, PANA, SSL, DTLS and IPv6-based security solutions employ the location of nodes. These security protocols secure communication channel between nodes rather than content. By adding security as a patch on IP, constraint-oriented IoT nodes perform their duties with more delays. Handling of mobile devices further complicates the situation. However, the IoT system is completely secured only when it ensures authentication, authorization, confidentiality, and integrity. 

Moreover, ICN offers security at the network layer because it provides contents-based communication. Content-based security provides easy and straightforward security to IoT contents without the involvement of third-parties or external intermediate nodes. Content-based security maintains the content integrity and data authentication. Furthermore, ICN contents can specify content access control towards its users because ICN contents are generally known as self-certified contents.    

\begin{figure*}[t]
\centering
\includegraphics[scale=.7]{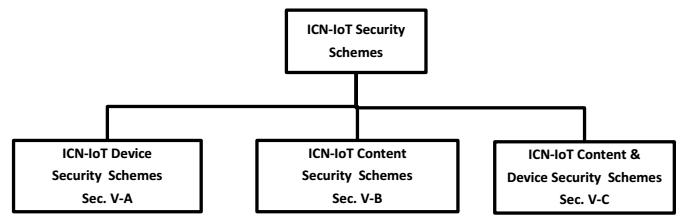}
\caption{ICN-IoT Security is Categorized into Three Categories: ICN-IoT Device Security Schemes, ICN-IoT Content Security Schemes and ICN-IoT Content and Device Security Schemes}  
\label{ICN-IoT-Security}
\end{figure*}

We categorize ICN-IoT (ICN-based IoT) security schemes into the following three categories: (i). ICN-IoT device security schemes, (ii). ICN-IoT content security schemes and (iii). ICN-IoT content and device security schemes.
ICN-IoT device security schemes deal with device authorization and authentication, and ICN-IoT content security schemes provide content integrity and confidentiality. Next, ICN-IoT content and device security schemes provide security to both content and devices. Categorization in ICN-based security for IoT can be visualized in Fig.~\ref{ICN-IoT-Security}.

\begin{table*}[t]
\centering
\renewcommand{\arraystretch}{1}
\label{Table-ICN-IoT-Sec}
\caption{ICN-IoT Security Schemes Are Summarized According to the Fig.~\ref{ICN-IoT-Security}.}
\begin{tabular}{|l|l|l|l|l|l|l|l|}
\hline
\toprule
\multicolumn{1}{c}{Ref.}    & \multicolumn{1}{c}{Model} & \multicolumn{1}{c}{\begin{tabular}[c]{@{}c@{}}Security\\  Perspective\end{tabular}}    & \multicolumn{1}{c}{Methodology}                                                                        & \multicolumn{1}{c}{Comparison}                                              & \multicolumn{1}{c}{\begin{tabular}[c]{@{}c@{}}Parameters \\ Evaluated\end{tabular}}             & \multicolumn{1}{c}{Finding(s)}                                                                                                                                                 & \begin{tabular}[c]{@{}l@{}}Simulator (OS,\\  Programming \\  Platform, \\  Language)\end{tabular} \\
\hline
\toprule
\multicolumn{8}{c}{ICN-IoT Device Security Schemes }
\\
\hline
\toprule
\cite{compagno2016onboardicng} & ICN & \begin{tabular}[c]{@{}l@{}}Device \\Authentication\\Device \\Authorization \end{tabular} &\begin{tabular}[c]{@{}l@{}}Symmetric\\ Cryptography\end{tabular} &\begin{tabular}[c]{@{}l@{}} ZigBee-IP\\ specification: \\EAP-PSK/ PANA \end{tabular}&\begin{tabular}[c]{@{}l@{}}Communication cost\\(communication\\ and computation)\\ and energy \\consumption (energy \\cost, memory cost)\end{tabular} &\begin{tabular}[c]{@{}l@{}} 87\% less communication\\66\% energy consumption\\ helps in confidentiality\\ of content, which in turn\\ maintain privacy\end{tabular} & \begin{tabular}[c]{@{}l@{}}ANALYTICAL \\EVALUATION\end{tabular} 
\\
\hline
\cite{mick2017laser} & NDN & \begin{tabular}[c]{@{}l@{}}Authentication\\Authorization\\Routing \end{tabular} &\begin{tabular}[c]{@{}l@{}}Symmetric\\ Cryptography\\and routing \end{tabular} & \begin{tabular}[c]{@{}l@{}}With its own\\ variants in terms\\ of increasing number\\ of nodes and distance \end{tabular} & \begin{tabular}[c]{@{}l@{}} Probability\\ Mass function,\\ Transmission burden,\\ convergence time \end{tabular} & \begin{tabular}[c]{@{}l@{}}Light weight\\ Authentication and\\ secure routing \end{tabular} & \begin{tabular}[c]{@{}l@{}}ndnSIM an\\ ns-3 extension\end{tabular}\\
\hline
\toprule
\multicolumn{8}{c}{ICN-IoT Content Security Schemes}
\\
\hline
\toprule
\cite{BoukAhmedKim2015} & CCN & Integrity & \begin{tabular}[c]{@{}l@{}}Base64 Format\\ on Content name\end{tabular} & No Comparison & No Implementation & \begin{tabular}[c]{@{}l@{}}Maintains\\ Integrity of \\content name\\ and device name\end{tabular} & \begin{tabular}[c]{@{}l@{}}Linux-based C++\\ programming\\ language\end{tabular}\\

\hline
%
\cite{arshad2017towards} & ICN & Integrity & \begin{tabular}[c]{@{}l@{}}SHA256 on\\ Content name\end{tabular} & No Comparison & No Implementation & \begin{tabular}[c]{@{}l@{}}Maintains\\ Integrity of \\content name\\ and device name\end{tabular} & \begin{tabular}[c]{@{}l@{}}No \\Implementation\end{tabular}\\
\hline
\toprule
\multicolumn{8}{c}{ICN-IoT Content and Device Security Schemes}
\\
\hline
\toprule
\cite{shang2014securing}& NDN &\begin{tabular}[c]{@{}l@{}}Data Privacy\\Data \\Authentication\end{tabular}&\begin{tabular}[c]{@{}l@{}}Data Privacy\\ through Access Control\\Data Authentication\\ through Digital\\ Signature \end{tabular} & \begin{tabular}[c]{@{}l@{}}No Comparison\end{tabular}&\begin{tabular}[c]{@{}l@{}}Analytically\\Evaluated\\Data Scalability\\preserved \end{tabular}&\begin{tabular}[c]{@{}l@{}} More responsive\\More scalable\\Less load as \\compared to IP-BMS \end{tabular}&\begin{tabular}[c]{@{}l@{}}Python-based\\ Application\\Data\\Visualization\\ Application\end{tabular}\\
\hline
\cite{enguehard2016slict}& ICN &\begin{tabular}[c]{@{}l@{}}Security and\\ geographic\\ forwarding \end{tabular}&\begin{tabular}[c]{@{}l@{}}Secure Beaconing\\ through \\OnboardICNg \end{tabular} & \begin{tabular}[c]{@{}l@{}}Vanilla ICN\\ forwarding\end{tabular}&\begin{tabular}[c]{@{}l@{}}No. of FIB entries,\\ energy cost,\\ Network overhead,\\ memory and\\ computation overhead \end{tabular}&\begin{tabular}[c]{@{}l@{}} OnboardICNg takes\\ extra computation,\\ energy and memory\end{tabular}&\begin{tabular}[c]{@{}l@{}}RIOT OS\end{tabular}\\
\hline
\cite{suarez2016secure}& CCN &\begin{tabular}[c]{@{}l@{}}Device\\ authentication\\Content Integrity \end{tabular}&\begin{tabular}[c]{@{}l@{}}PK Cryptographic \\Suite Symmetric\\
Encryption using AES \end{tabular} & \begin{tabular}[c]{@{}l@{}}Arduino board\\ for proof of \\concept\end{tabular}&\begin{tabular}[c]{@{}l@{}}1.Info. Freshness\\level, 2.Interest\\Range stability 3.\\Energy consumption \\with or without\\security feature via\\UDP and CCN 4. \\Packet overhead \\estimation\end{tabular}&\begin{tabular}[c]{@{}l@{}}1. Avg. Service\\time is stable for\\interest rate less\\than 24 request/s 2.\\Energy Consumption\\with security feature\\ 0.33\% Without security\\with CCN feature 0.28\%\end{tabular}&\begin{tabular}[c]{@{}l@{}}ndnSIM 1.0\end{tabular}\\
\hline
\cite{sicari2017secure} & ICN &\begin{tabular}[c]{@{}l@{}}Privacy, trust,\\content integrity,\\ confidentiality,\\ authentication,\\ access control \end{tabular}&\begin{tabular}[c]{@{}l@{}}device discovery\\ service discovery\\ secure subscription,\\ Secure naming service,\\Secure content delivery\end{tabular}&\begin{tabular}[c]{@{}l@{}}No Implementation\end{tabular}&\begin{tabular}[c]{@{}l@{}}No Implementations\end{tabular}&\begin{tabular}[c]{@{}l@{}}Secure ICN-IoT\\ Architecture\end{tabular}&\begin{tabular}[c]{@{}l@{}}UML diagrams\end{tabular}\\
\hline

\end{tabular}
\end{table*}
\subsection{ICN-IoT Device Security Schemes}
In \cite{compagno2016onboardicng}, an ICN-based secure protocol is proposed which provides security regarding both the authentication and authorization of IoT devices. They call this ICN-based security protocol as on-boarding protocol (OnboardICNg). OnboardICNg protocol authenticates every joining device and authorizes it through authorizing this device. They consider authentication and authorization manager (AAM) for initial key sharing. Key is shared between new joining device and AAM to guarantee it as a secure IoT device. The new device knows the naming format of publishing and requesting any content. A single key is supposed/assumed to provide authentication, integrity and confidentiality. They used and modified, authenticated key exchanged protocol (AKEP2) according to the ICN design for IoTs. Through OnboardICNg, IoT network is secured from internal and outside adversaries. They compare OnboardICNg with Pre-Shared Key Extensible Authentication Protocol (EAP-PSK)/PANA in terms of communication cost (both communication and computation costs) and energy cost (both energy and memory costs). They find OnboardICNg is more useful for IoTs with 87\% and 66\% reduction in communication and energy costs respectively as compared to EAP-PSK/PANA. However, authors do not provide any simulations and present only analytical results for the proposed protocol.     
 
 Authors in \cite{mick2017laser} enhances Onboarding authentication protocol and combines routing with it. They call the proposed protocol lightweight authentication and secure routing (LASeR) protocol. They consider IoT smart cities made through islands. The considered scenario has anchor nodes, standard nodes and gateway nodes. Among which only standard nodes are IoT nodes.  An island manager (IM) just like AAM in \cite{compagno2016onboardicng} is used to authenticate and authorize the nodes. LASeR protocol works in three steps: discovery phase, authentication phase and advertisement phase. They evaluated LASeR in terms of convergence time and transmission burden for the different number of nodes and increasing distances among nodes. The LASeR only focuses on authentication with routing. However, IoT nodes do not involve in this whole procedure, and they delegate their duties to anchor nodes and IM. Like \cite{compagno2016onboardicng} they also talk about securing the IoT applications and nodes as a whole.
  
 \subsection{ICN-IoT Content Security Schemes}
 In \cite{BoukAhmedKim2015} authors have presented secured content naming scheme where a content name is secured using Base64 Format. This work only considers multimedia contents fetched by vehicles. The secured part is included at the end of Interest packet and can be calculated by taking the hash of attributes of content or a public key of the vehicle. They have programmed it in Linux-based C++ programming. They have only considered vehicles and not static devices. 
 
 In \cite{arshad2017towards} we propose an IoT content naming scheme. IoT applications categorization is updated and a universal hybrid naming scheme is proposed. Content is secured using SHA256 to maintain integrity. Fetched content name and its sub-type name is encrypted through SHA256. Moreover, the name of the node which is originating the Interest is also encrypted through SHA256. Security is preserved in the context of integrity. However, no implementation results are presented.
\subsection{ICN-IoT Content and Device Security Schemes}
 NDN-based architecture to secure any building is presented in \cite{shang2014securing} and installed in University of California at Los Angeles (UCLA). It is a prototype to show the performance achieved by NDN instead of IP-based security systems. Their proposal consists of three main entities, end users, gateway and a manager application. Gateway and sensor devices run IP-based building management system (BMS) protocols. Manager application is controlled by a human operator who authorizes out of band users. Manager application is also responsible for NDN management and auto-configuration of sensors and gateways. Gateways publish contents into NDN repositories. NDN repositories are responsible for responding user queries about sensors data. In NDN-based BMS, they followed and designed hierarchical naming to name devices and contents. They used public keys of the user and append it as the last component of content name by calculating its hash through SHA256. Two lists are maintained to maintain user privileges. Each gateway has an access control list (ACL), which is a list of identities of authorized users. Another list, access privilege list (APL) which is maintained by every user of BMS, contains the data namespaces that any user can access. APL is also published in NDN repositories. To provide the mapping between content namespaces and user IDs both lists (i.e., ACL and APL) are responsible which saves BMS manager from traversing entire BMS application to update user privileges. Both ACL and APL can be published as NDN data. They consider capability-based access control. ACL lists the capabilities to access sensor data, and the user gets capability-certificate to access data. During gateway configuration, NDN packets are signed and encrypted using the symmetric key to secure from man-in-middle attacks. Sensor data is encrypted through the shared symmetric key to providing access-control and published in JSON format. Gateway generates and distributes symmetric keys while going through ACL. It publishes encrypted key (encrypted through the user's public key) asymmetrically. Data packet also contains time-stamp of the decryption key to ensure content-based security. Python-based data publishing service is used to publish data and browser-based data visualization application. The data publishing service packs data in JSON format into NDN repositories. The user issues interests using data visualization application and can employ time-stamp filter. The user gets encrypted data and decryption is performed through the encrypted symmetric key. Data is encrypted using AES-CBC cipher. The BMS system presented in this asynchronous approach is not suitable for IoT a situation where fresh data is required from a sensor because of sensor uploads data into NDN repositories first. However, it enables caching, lowers load on the data server and preserves IoT scalability as data is secured via encryption only single time.   

In \cite{enguehard2016slict} authors discuss forwarding and security for ICN-based IoT. Geographic forwarding is implemented due to its low control traffic for sending data towards the destination. It involves the location of destination for content transmission and thus lower network resources usage while maximizing energy life of IoT devices. To provide security, authors force the use of symmetric cryptography through OnboardICNg. They state that OnboardICNg authenticates locally two nodes and verifies that both are parts of a trusted network. Through provided shared symmetric key, nodes authenticate each other to build a secured network. Next, they discuss secure push mode through secure beaconing. Insecure beaconing can introduce DoS and wormhole attacks. Through broadcasted shared symmetric keys, sensors distinguish the beacons from the trusted users. Beacon messages are secured by encrypting these through the broadcast keys provided by OnboardICNg. Further messages after the beacon, contain MACs generated through encryption using broadcast keys. However, if the neighboring node is tempered then this scheme is not resilient.
They evaluate their proposal in RIOT OS in terms of computation, network and memory footprints. It takes 28 to 35 extra bytes per message like beacon, interest and data message during transmission in 802.15.4-based OpenMote. AES-CCM takes more energy both in software and hardware, and it is one order lower than the transmission of messages. Cost of memory footprints includes three keys per node and authors state that this is likely a negligible space available on most recent boards like OpenMote. However, the main aim of this proposal is to evaluate geographic forwarding in ICN-based IoT. They also evaluate OnboardICNg on both hardware and software and find that security comes at a cost. This proposal secures ICN-based IoTs through securing IoT devices and contents.
  
  In \cite{suarez2016secure}, authors discuss benefits and challenges of applying ICN for IoT. They consider two content requests, (i) when any user wants an action performed by any device and (ii) when the user requests the current content of the device. Their proposal consists of a gateway, admin, clients with the same name-space, IoT devices and other clients. Gateway is the central device which connects with admin, IoT devices and clients to provide interoperability between powerful and constraint-oriented devices. This gateway is also placed to cope with heterogeneous devices differentiated as devices from different name-spaces. Gateway exchange management-content-information with IoT devices through the reference point Mdg. This Mdg as a reference point is responsible for secure content centric communication with IoT devices. Client and gateway mutually authenticate the security mechanism for full proof content exchange in CCN. Through the discovery procedure, the client discovers a list of IoT devices. In its working, as step 1, the client first expresses an interest in the form of CCN name. In step 2, the gateway receives this interest and respond with the data packet. Data packet indicates content protection and also provide information to the client for encryption algorithm and key sizes. For normal CCN phenomenon, data also incorporate shorthand identifier for the gateway (i.e., GW publisher ID). GW publisher ID is calculated through a cryptographic digest of its public key and key locator is responsible for the actual location of the public key. In step 3, in order to get appropriate key, the client issues an interest in the protection of exchange information. Then, the client gets verified through the gateway to enable IoT service routine. When the client is authenticated, the gateway generates a random systematic key SKcg (128 bit AES key) for cryptographic functions. This SKcg key along with its related information is encrypted with the public key of the client as extracted from the data packet in step 4.  Data provided by the gateway is verified and decrypted by the client through its SKcg. The client also generates a Message Authentication Code (MAC) over the whole interest by using the session key SKcg. In step 5, MAC and a unique nonce value are appended with CCN name to prevent malicious attacks. Gateway verifies nonce and MAC component and replies to interest message with the data packet in step 6. As step 7, the client can retrieve information from the gateway by issuing interest after validation of client. Then gateway reply client in accordance with client specific policy in step 8. The gateway-based proposed design presented in this paper have the flexibility to adopt according to environment and organization. It also enables security feature through the built-in support of automatic discovery and registration process which is the uniqueness of this design. It also reduces the overall incoming interest packets. The result shows that the average service time of interests is stable for 25 requests per second. This work is suitable for IoT as it can scale up with less overhead and secures both IoT contents and devices.
  
  In \cite{sicari2017secure} authors proposed an ICN-based secure architecture for IoT. The proposed ICN-IoT secure architecture provides a trust model for nodes and links, privacy for sensitive information and effective access control system. Five components including IoT nodes (Content producers), service consumers, ICN-IoT server, local server gateway (LSG) and aggregator, build proposed ICN-IoT middleware. They integrate security with ICN-IoT architecture \cite{zhang2013icn}
interactions involving, device discovery, service discovery, naming service, user registration, and content delivery. Authentication of devices is performed through device discovery phase. Secure device discovery is ensured when any new device joining IoT network sends its device ID, signature key and certificate; this triplet is sent towards aggregator where it verifies and stores this new device information. Then aggregator issues an action key encrypted through the signature key. If the new joining device is not a certified device, then it can send its device ID only. In this case, the aggregator can issue a signature key and certificate. This method can be helpful for mobile devices authentication. 

Further service discovery is used by IoT users to get any service. IoT user connects with ICN-IoT server through sharing its both signature key and device ID. Upon successful access grant, the user further sends its actual query/request in encrypted form through its action key and signature key. ICN-IoT server forwards this request towards aggregator. Aggregator decrypts and satisfies the request with the help of IoT nodes and sends relevant response towards corresponding IoT user. Secure naming service provides security to names of IoT devices. Aggregator sends device ID, a signature key and action keys towards LSG which in turn assigns the name to device and replies name to the aggregator. Aggregator sends device name towards device by encrypting it through action key of the device. During a subscription, a user needs a secure subscription which is performed through secure user registration. User contacts ICN-IoT server by sharing its own information along with device name. ICN-IoT server replies user with ID, signature key and password (which user can change). Secure content delivery from the device is ensured by sending device name, ID encrypted with signature key and data encrypted with action key to the aggregator. Aggregator decrypts data and sends to ICN-IoT server. ICN-IoT server again encrypts data with the action key of the user and sends towards the user. Proposed ICN-IoT architecture aims to secure both content and device by maintaining privacy, authentication, confidentiality, and integrity. However, the authors did not provide simulations to verify the results. They only provide UML diagrams to describe their proposal.
  \subsection{Summary and Insights}
In this section, we have surveyed ICN-based security schemes in terms of IoT and classified these security approaches into three categories. In the first category, we listed and summarized those approaches which handle the ICN-based security of IoT devices. These approaches mainly provide authentication and authorization of IoT devices. The second category, ICN-IoT content-based security schemes mainly deal with content and aimed to provide content integrity, non-repudiation and confidentiality. The resulting contents are self-certified which can specify its owner details and content details. In the third category, ICN-IoT content and device security schemes, those approaches are discussed which include both device and content properties. ICN security approaches in this class mainly focus on securing the whole IoT system while providing content integrity, confidentiality and device authentication and authorization. Moreover, some techniques also consider access-control-management which aims to specify the list of intended users.

Our survey finds that ICN-based security schemes must be designed which involve IoT environment characteristics; for example, constraint-oriented nature of IoT devices. As IoT applications can involve push operations; for instance, an actuator IoT device can only perform a simple action like turning some devices on/off if this query/command is received from authenticated and trusted IoT node. However, most methods which are discussed above, apply security methods over interest and data messages. Therefore, there is a need to ensure that security mechanisms must provide authenticated requests along with push support enabled. 

Moreover, public key cryptography (asymmetric cryptography) cannot be implemented for IoT resource-constraint (i.e., in terms of memory and processing) devices because of its resource-intensive nature. ICN-IoT content security schemes which embed security information at the end of query/interest packets as last named component, result in lengthy request packets and increase complexity to be processed by IoT constraint-oriented nodes. For this reason, lightweight security solutions to maintain confidentiality, integrity and authentication are optimal and feasible choices for IoT constraint-oriented nature. 

From this perspective, symmetric key cryptography can play an important part and is explored in many approaches like \cite{shang2014securing}-\cite{burke2014secure}-\cite{compagno2016onboardicng}-\cite{mick2017laser}. As symmetric cryptography approaches need to maintain keys and exchange of these keys is required before any communication. However, these pre-shared keys cause extra overhead and also make symmetric key cryptography inflexible for IoT.

Besides these, nowadays Elliptic Curve Cryptography (ECC) is being explored for IoT constraint-oriented devices because of its simplicity and extra lightweight nature. ECC utilizes elliptic curve theory to produce better cryptographic keys in terms of size and efficiency. As compared to the RSA algorithm, where the keys are generated from the product of two large prime numbers, ECC creates them through the properties of elliptic curve equation. It relies on the difficulty of solving the elliptic curve discrete logarithmic problem. Although the key size in ECC is smaller, it can provide as good security as any other traditional method such as RSA  and eventually reduces the processing cost. Therefore, it is expected from ECC to provide essential security features for secured ICN-based IoT.  

Finally, to conclude, our survey of ICN-IoT security schemes indicates that no single solution fulfils all requirements of IoT nodes and applications. Therefore, ICN-based IoT security solutions must be designed in a flexible way which includes both IoT application requirements and device specifications and capabilities.
\section{\sobia{ICN-IoT Mobility Schemes}}
IoT networks can include hybrid and heterogeneous devices in terms of mobile and non-mobile (i.e., static) devices. While most of the IoT applications such as smart home, smart grid, smart building require mostly static devices. However, other applications like smart transport, smart vehicles, smart mobile networks involve more mobile devices as compared to static devices. Therefore, mobile devices are an important part of IoT, and thus their management also becomes essential.

Although there are other mobility models (like nomadic and pervasive), however in IoTs, cellular mobility model plays an important role. In cellular mobility, wireless networks are divided into cells and each cell has specific radius and area of service. When mobile devices move from one cell to the next, they face a situation called a handoff condition. Therefore, handoff-management also becomes an essential factor to solve.

In ICN-based IoTs, both subscriber and producer can be mobile devices. As described and discussed before, ICN-IoT mobile subscriber can benefit from connection-less and receiver-driven nature of ICN. Therefore, in this way, the mobile subscriber can re-issue interests for which they did not receive data. To support mobility, DTN function does not need heavy protocols like Mobile-IP. In contrast, publisher mobility is complex to manage as it requires some additional operations.

We categorized ICN-based mobility schemes into ICN-based IoT producer mobility management schemes and other ICN-IoT Mobility schemes. In the first category, those schemes are combined, in which ICN-based producer mobility is discussed. ICN producer mobility scheme further categorized into anchor-less producer mobility. In other producer mobility schemes, ICN-IoT smart forwarding schemes are discussed.  

\begin{table*}[]
\centering
\caption{\textbf{ICN-IoT Mobility Schemes}}
\label{Table-ICN-IoT-Mobility}
\begin{tabular}{|c|c|c|c|l|l|c|c|}
\hline
\toprule
\multicolumn{1}{c}{Ref.}    & \multicolumn{1}{c}{Model} & \multicolumn{1}{c}{\begin{tabular}[c]{@{}c@{}}Mobility\\  Perspective\end{tabular}}    & \multicolumn{1}{c}{Methodology}                                                                        & \multicolumn{1}{c}{Comparison}                                              & \multicolumn{1}{c}{\begin{tabular}[c]{@{}c@{}}Parameters \\ Evaluated\end{tabular}}             & \multicolumn{1}{c}{Finding(s)}                                                                                                                                                 & \begin{tabular}[c]{@{}l@{}}Simulator (OS,\\  Programming \\  Platform, \\  Language)\end{tabular} \\ \hline
\toprule
\multicolumn{8}{|c|}{ICN-IoT Producer Mobility}   \\ \hline 
\toprule
\cite{anastasiades2014information} & NDN & \begin{tabular}[c]{@{}l@{}}Producer\\ Mobility \end{tabular} & \begin{tabular}[c]{@{}l@{}}Content transfer\\content discovery \end{tabular} & No Comparison & No Implementation & \begin{tabular}[c]{@{}l@{}}Delegating content\\retrieval to agents\\ is better\end{tabular} & \begin{tabular}[c]{@{}l@{}}CCNx \end{tabular}\\
\hline
\cite{meddeb2017producer} & NDN & \begin{tabular}[c]{@{}l@{}}Producer\\ Mobility \end{tabular} & \begin{tabular}[c]{@{}l@{}}Survey \end{tabular} & No Comparison & \begin{tabular}[c]{@{}l@{}}delivery cost\\path length\\interest routing\end{tabular} & \begin{tabular}[c]{@{}l@{}}Name-based routing\\ is better\end{tabular} & \begin{tabular}[c]{@{}l@{}}Analytical \\Evaluation\end{tabular}\\
\hline
\cite{zhang2016survey} & NDN & \begin{tabular}[c]{@{}l@{}}Producer\\ Mobility \end{tabular} & \begin{tabular}[c]{@{}l@{}}Survey \end{tabular} & No Comparison & \begin{tabular}[c]{@{}l@{}}Signal overhead\\security\\name-changes\\dependency on RV\end{tabular}& \begin{tabular}[c]{@{}l@{}}data depot+tracing,\\data depot+mapping\\are better\end{tabular} & \begin{tabular}[c]{@{}l@{}}Analytical \\Evaluation\end{tabular}\\
\hline
\toprule
\multicolumn{8}{|c|}{Anchor-less ICN-IoT Producer Mobility}   \\ \hline 
\toprule
\cite{auge2015anchor}-\cite{auge2016map}                                                            & \begin{tabular}[c]{@{}c@{}}NDN\end{tabular}                            & \begin{tabular}[c]{@{}c@{}} Producer\\ Mobility \end{tabular} &\begin{tabular}[c]{@{}c@{}} IU and IN \\through \\Sequence \\Numbers \end{tabular}                                                                 & \multicolumn{1}{c|}{GR, AB, TB}                                                                                                                                  & \begin{tabular}[c]{@{}l@{}}Avg. Packet loss,\\ delay \& hop-count \\No. of messages,\\ signaling overhead\\link utilization\end{tabular} & \begin{tabular}[c]{@{}c@{}}Better network cost\\ \& user performance\end{tabular}                                                                   & \begin{tabular}[c]{@{}c@{}}ndnSIM\end{tabular}                     \\ \hline
\cite{compagno2017secure}                                                            & \begin{tabular}[c]{@{}c@{}}NDN\end{tabular}                            & \begin{tabular}[c]{@{}c@{}} Secure\\Producer\\ Mobility \end{tabular} &\begin{tabular}[c]{@{}c@{}} Hash and \\Hash chains \end{tabular}                                                                 & \begin{tabular}[c]{@{}c@{}} MD-1,-5, SHA256,\\ DSA, RSA\end{tabular}                                                                 & \begin{tabular}[c]{@{}l@{}}Computation\\ Overhead \\Storage overhead \end{tabular} & \begin{tabular}[c]{@{}c@{}}Lightweight attestation\\ \& Scalable\end{tabular}                                                                   & \begin{tabular}[c]{@{}c@{}}ndnSIM\end{tabular}                     \\ \hline
\end{tabular}
\end{table*}
\subsection{ICN-IoT Producer Mobility}
Producer mobility is accomplished in two steps. Firstly, producer location is needed to find and trach along with easy session maintenance. Then it is identified that the architecture is coupled or decoupled in terms of name-resolution and data-transfer. In coupled architecture, producer advertises content prefix from its new location. While in decoupled approaches, resolution information is needed to update from the new location.

In \cite{anastasiades2014information} producer mobility support mechanisms and their disadvantages are discussed in three categories. Routing-based producer mobility is provided by updating the routing tables which involve the forwarding of information queries. However, the routing-based approach is not suitable to provide scalability
of routing tables. Second, the indirection approach requires some
extra nodes (home-agents) which keep track of nodes locations and forward interests to the updated location of the mobile producer. Drop-acts of this approach lies in the form of extra management of content names and their name-resolution (i.e., information of producers). Also, every query and data message also visit this home-agent. The third approach, resolution-based include content updated location (or information about updated location) in data message as a response of user query. Resolution based approach incurs the overhead of this one extra packet. They further discuss both content discovery and transfer mechanisms. This work discusses the feasibility of ICN mobility in terms of both mobile producers and consumers in opportunistic and mobile
networks which is a definite part of ICN-IoT. 

In \cite{meddeb2017producer} NDN-based producer mobility is discussed for IoT. They discussed NDN-based producer mobility support through four approaches. The first approach solves producer mobility by
utilizing the location information through the location resolution
system (LRS). Producer updates LRS about its location after
moving. LRS keeps the record of content name prefix and its
corresponding producer. Consumer requests the location of the content producer by sending the message having content prefix
towards LRS. In the second triangular approach, interest message
is sent towards the previous location. Then using FIB update, it is rerouted towards the new location. The data message is delivered firstly
towards old location and then from there, it is forwarded to the consumer. In the third locator/identifier separation approach, every
content is managed in two parts by its producer. Content first
part is its identifier and the second is its locator. In identifier, prefix
or content name is stored and in the locator, the location of the router
(to which it is currently connected) is saved. After producer
mobility, it changes its locator value with the location of new
connected router. The fourth approach, routing-based approach finds the data through the name-based routing protocol. Name-based routing protocol tries to find the cached copies of data towards the path of the original producer. Name-based routing can be implemented through decentralized routing using flooding and distance-based greedy routing protocol. Thus its complexity depends on the routing protocol. They expect that name-based routing scheme can perform better in IoT due to its average cost for packet delivery, less handover latency and optimal routing patch length. However, they did not propose any technique for NDN-IoT producer mobility.

In \cite{zhang2016survey}, authors survey producer mobility and categorized into four categories: (i) mobile producer (MP) mapping, (ii) MP tracing (iii) data depot and (iv) data spot. In MP mapping, MP informs rendezvous (RV) node about its point of attachment (PoA) and data can be obtained through mapping provided by RV or RV tunnels the interest messages towards MP. In MP tracing, interest messages can use traces of MP on the way towards RV and get forwarded towards MP without involving RV. In the data depot, a fixed location saves the data produced by MPs and can forward the data in response to interests with involving MP in this whole procedure. Finally, in the data spot, new MPs generate data in order to fulfil the interest. However, in IoT, data depot along with MP tracing (or mapping) plays the part due to nature of IoT applications. Moreover, data depot along with tracing can enhance interest satisfaction rate as IoT devices may run out of battery more often and traces can provide a direct path towards MP.  
\subsubsection{Anchor-less Producer Mobility}
In \cite{auge2015anchor}, proposed producer mobility management (MM) scheme is designed to meet 5G requirements of low latency, low network overhead and overall fast speed. MM schemes are categorized into three classes: (i) anchor-based, (ii) anchor-less and (iii) rendezvous-based. In anchor-less MM, any node is responsible for providing information about its new location. In rendezvous-based MM, dedicated nodes are responsible for providing resolution of identifiers into locators. In anchor-based approach, a specified node is responsible for all nodes movements and direct messages to the new locations of moved nodes. They have proposed the anchor-less MM system to support delay-sensitive applications like smart health. When a patient is moving and acts as a mobile producer, its fast MM is important. They used stateful forwarding, ICN in-network caching and defined forwarding mechanism, to update and populate Temporary FIB (TFIB) from producer new location towards its former location. MM does not need global routing updates and any change in the content name. It employs the distributed and dynamic ICN forwarding and eliminates the need for in-network anchors while limiting the MM towards edge nodes. Anchor-less MM is lightweight in nature because it limits signaling and maintains temporary change or state by in-network nodes. To support latency-sensitive transmissions during high mobility, network notifications and discovery methods provide necessary support. Anchor-less producer mobility is ensured in three simple following steps. Every mobile producer updates content (it produces) as a list of prefixes to its new PoA after establishing a link with this PoA in a defined message called Interest Update (IU). After a relocation, producer changes router and populates TFIB using forwarding update operation. Consumer interest is forwarded towards producer using this TFIB information or using FIB along with discovery mechanism. In \cite{auge2016map}, they have evaluated their proposed anchor-less producer MM and called it Map-Me. For delay-sensitive applications, producer left its traces on the way to its new location and they named it Interest Notification (IN). Due to its lightweight nature, IN supports delay-sensitive applications. They also provide both analytical and simulation evaluation. Simulation is carried in ndnSIM with total 36 wifi nodes. They found proposed MM better than global routing, tracing-based and anchor-based approaches in terms of average packet loss, average packet delay, average hop counts, number of messages, signaling overhead and link utilization. This anchor-less MM is highly suitable for IoT applications and delays sensitive applications like smart health. 

In \cite{compagno2017secure}, authors identified loop-holes of \cite{auge2016map} and propose a prefix attestation protocol to secure trace-based producer mobility. Protocol Map-Me can be compromised when IU came from an attacker. It can pollute cache and disturb the privacy of consumers and edge routers. Session-key and signature-based used for securing routers. However, both are not suitable for 5G networks. In their prefix attestation protocol, the producer sends minimal security context towards the registration server to generate valid IU. This security context is distributed locally among local routers and they use this information to validate IU locally. Security is maintained while allowing fast validation and generation of valid IUs through hash functions and hash chains, respectively. They evaluate the attestation protocol analytically in terms of goodput. Goodput decreases when because IUs take resources. Hash chains maintain optimal goodput in case of one hash or multiple hashes per IU verification. Around 50 MB is required for millions of mobile users in one router and proposed prefix attestation protocol is thus more scalable.
\subsection{Other approaches in ICN-IoT Mobility}
In \cite{AhmedBoukKim2015} a forwarding mechanism is presented for vehicles by incorporating one immediate vehicle resources. It ranks the vehicle upon multiple factors and selects one as forwarder among all vehicles.
However, it does not account for provider mobility (i.e., adhesive issue of ICN mobility). Moreover, in\cite{BoukAhmedYaqubEtAl2016}, authors provide a scheme DPEL (Dynamic PIT Entry Lifetime) to reduce the number of PIT entries. Therefore, it minimizes the usage of the battery of mobile nodes and makes routing and forwarding easy and fast. 
\subsection{Summary and Insights}
This section presented ICN-based IoT mobility and categorized presented schemes. As ICN supports consumer mobility naturally, but mobile producer support is undefined. ICN consumer can re-issue interest for any missed packet and can get data after location change. ICN producer mobility is hard to handle.

IoT needs fast data continuity in real-time applications. Moreover, resource-constrained nature of IoT devices put more challenges like tracking mobile devices in terms of old and new locations of mobile devices, reducing handover delay and simplify mobility management and handling with less number of packets. Therefore, in this context, anchor-less producer MM \cite{auge2015anchor}-\cite{auge2016map} can be employed for IoT environment and can be secured further through the hash chains method presented in \cite{compagno2017secure}. 

Moreover, in other ICN-IoT schemes, those schemes are included which try to make IoT mobile node lighter while minimizing PIT entries and selects the best forwarder among available vehicles. 

However, there is not any single solution exists for ICN-IoT producer mobility and handoff management. This may be because IoT general applications like smart home involve mostly static devices. Therefore, mobility is the most ignored perspective and available as a fruitful research direction. 
\section{ICN-IoT Operating Systems and Simulation Tools}
There are a lot of IoT Operating Systems (OS) and simulation tools which can be used for ICN-IoT. In \cite{HahmBaccelliPetersenEtAl2015} famous IoT OSs including Contiki \cite{DunkelsGronvallVoigt2004}, FreeRTOS \cite{freertos}, RIOT \cite{BaccelliHahmGunesEtAl2013}, TinyOS \cite{levis2012experiences} and OpenWSN \cite{open-wsn} are presented under the category of open-source and closed-source (which are not available commercially). \sob{Among these, we only discuss which can be used for both IoT as well as ICN implementations.} On the other hand, specific ICN simulators (ndnSim \cite{ndnsim}, ccnSim\cite{ChiocchettiRossiRossini2013} and Icarus \cite{SainoPsarasPavlou2014}) are presented in \cite{tortelli2014ccn}. However, from this paper perspective, it can be seen in Table~\ref{Table-ICN-IoT-OS-Sim} that ndnSIM for NDN is the most explored simulator for ICN-IoT.
\begin{table*}[]
\centering
\caption{ICN-IoT OS and Simulation Tools Analysis}
\label{Table-ICN-IoT-OS-Sim}
\begin{tabular}{|c|c|c|c|}
\hline 
\rule[-1ex]{0pt}{2.5ex} • & ndnSIM \cite{ndnsim} & Contiki OS/Cooja Simulator \cite{HahmBaccelliPetersenEtAl2015,DunkelsGronvallVoigt2004} & RIOT OS \cite{BaccelliHahmGunesEtAl2013} \\ 
\hline 
\rule[-1ex]{0pt}{2.5ex} Ref. \# & \cite{QuevedoCorujoAguiar2014a}-\cite{QuevedoCorujoAguiar2014}-\cite{HailAmadeoMolinaroEtAl2015}-\cite{HailAmadeoMolinaroEtAl2015a}-\cite{quan2014social}-\cite{mick2017laser}-\cite{auge2016map} & \cite{AbidySaadallahyLahmadiEtAl2014}-\cite{CCNxsaadallah2012}-\cite{sobia-iot2018} & \cite{BaccelliMehlisHahmEtAl2014}-\cite{enguehard2016slict} \\ 
\hline 
\rule[-1ex]{0pt}{2.5ex} Total \# of  Ref.  & 7 & 3 & 2 \\ 
\hline 
\end{tabular}
\end{table*} 
\subsection{Contiki OS with Cooja Simulator}
Contiki \cite{HahmBaccelliPetersenEtAl2015,DunkelsGronvallVoigt2004} is an open source and flexible operating system developed at the Swedish Institute of Computer Science (SICS) in Sweden. It is a very lightweight operating system for sensor nodes which are severely resource-constrained regarding power, memory, processing power and communication bandwidth. Contiki is developed in C language and is event driven. Main features of Contiki operating system include the support of preemptive multithreading per-process and dynamic loading and unloading of code at runtime. A Contiki configuration consumes 40-kilobytes of ROM and 2-kilobytes of RAM. The communication between different processes always goes only using the kernel of the operating system. A full installation of Contiki operating system includes many features such as preemptive multithreading, TCP/IP networking, proto-threads, Graphical User Interface, multitasking kernel, IPv6, web browser, simple telnet client, personal web server, and virtual network computing. Its current version is 3.0 released on August 26, 2015. 
Cooja Simulator \cite{cooja} is the Contiki network simulator. Cooja allows large and small networks of Contiki motes to be simulated. Motes can be emulated at the hardware level, which is slower but allows precise inspection of the system behavior, or at a less detailed level, which is faster and allows simulation of larger networks. Contiki along with Cooja Simulator makes it a perfect combination for ICN-IoT related research.
\subsection{RIOT OS}
RIOT \cite{BaccelliHahmGunesEtAl2013} is licensed as LGPL (Lesser General Public License) and an open-source operating system for sensor nodes in the Internet of Things. RIOT OS is a microkernel-based operating system inherited from Fire Kernel \cite{WillSchleiserSchiller2009} which matches the various software requirements for IoT devices. The key design objectives for RIOT OS include energy-efficiency, small memory footprint, modularity, and a developer-friendly programming interface, which make RIOT the best choice to power the widest spectrum of IoT devices. Implementation and design of RIOT have the ability to deal with the various challenges in powering of constrained devices networks. RIOT also provides both real-time capabilities and full multi-threading. RIOT provides the C and C++ programming language supports for applications.
\subsection{Other Simulators}
NDN architecture can be simulated using its own specific ndnSimSimulator. This ndnSim \cite{ndnsim} is an NS3-based simulator and provide simulation for NDN and CCN. 

Mini-CCNx \cite{ccnx} is a tool for agile prototyping of ICN-based on the CCN model. Mini-CCNx is used to build several CCN topologies, each with hundreds of nodes, great agility, and flexibility. These topologies can be run directly on laptop/desktop, in a local VM or cloud. Moreover, the best is: the code which user runs on Mini-CCNx is the same code used in a real network. This feature adds a realistic behavior to simulation tests. Each Mini-CCNx node (host or router) runs the official Project CCNx's; therefore, the user uses the official CCN implementation.

ICN Simulator, the Information-Centric Network Simulator which is developed by the University of Essex works with OMNET++ simulation environment. It provides PURSUIT architecture functionalities. It is able to simulate a large number of nodes and publisher-subscriber pairs and produce a massive amount of information, providing an insight on the new techniques introduced in the topology management of the information-centric network.

Icarus \cite{SainoPsarasPavlou2014} is a caching simulator which supports multiple caching schemes and replacement schemes. It is a Python-based general tool to evaluate and implement ICN caching schemes. It does not support any specific ICN flavor but a simple environment to work with ICN caching.

\section{Issues, Challenges, and Future Research Directions for ICN-IoTs}
In this section, we present issues with the current solutions for ICN-IoTs and identify future research directions that need to be solved by the research community.
\subsection{Naming}
Most of the ICN-based IoT naming research is conducted for CCN/NDN hierarchical naming. As CCN header is of fixed size (8 bytes) \cite{ccnpacketformat}. Therefore, to apply CCNx (with fixed header) for IoT low-power and constraint-oriented devices, header compression techniques can be explored to support small data packets.

However, NDN packet \cite{ndnProject}-\cite{ndnpacketformat} does not have fixed length header. For small data packets (like mostly IoT applications have short length data to transmit in a response of a query or to send command towards any sensor or to just acknowledge the command to or to send current state of any sensor), NDN packet formats with variable length headers provide good support for IoTs applications \cite{ndnpacketformat}. 

In addition, as CCN/NDN naming follows the hierarchical structure that generates long and variable length names, and these long names can be utilized to build applications that have to update their status (or sensor values) continuously. For instance, heart-beat of a specific person having any sort of cardiac disease. This can help the doctor to fetch the heartbeat value of that patient recorded at any specific time instant. Conversely, long names raise the problems to fit in Zigbee maximum payload of 127 bytes, so naming schemes consider this factor also. Additionally, hierarchical names are human-readable, thus, still, there is a need to design secured hierarchical compact naming scheme to provide original data in the case of privacy-sensitive applications like smart-health. Furthermore, in this context, the work in \cite{kietzmann2017need} analyses the aspects of layer two communication in an NDN-based IoT. Findings indicate that L2 broadcasting has a severe negative impact on efficiency and reliability of content replication, which can be mitigated using a proper name-to-MAC-address mapping. Hence communication to groups should a layer three control and take advantage of the address mapping. Moreover, in \cite{gundogan2017information} authors provide a system (i.e., that translate NDN names and MQTT topics) to show how these elements can be assembled to build a safety-critical surveillance environment for the IoT.

Moreover, lookup for length-varying names is expected to be complex. Therefore, it is quite stimulating and difficult to design such a lookup system for IoT constraint-oriented devices\cite{zhang2013icn}-\cite{icn-ietdraft02}.
   
Current literature investigated and proposed naming scheme for any single application, for instance in \cite{AmadeoCampoloIeraEtAl2015} and \cite{BoukAhmedKim2014} ICN naming schemes are proposed for smart-home and VANETs respectively. Therefore, we stimulate ICN-IoT research community to put efforts to find and develop a naming scheme with carefully selected general, collective and public prefixes to cover (identify) and refer all IoT applications \cite{arshad2017towards}-\cite{sobia-iot2018}. We are still looking for a general and appropriate naming scheme that can solve all identified constraints.     
\subsection{In-Network Caching}
Though identified as the major beneficial feature of ICN for IoTs, ICN-IoT caching has received a lot of attention from the research community. By employing ICN caching in IoTs can save network bandwidth, reduce latency to get data and improve the battery life of IoT devices \cite{AmadeoCampoloMolinaro2014}. 

Mostly ICN-based caching schemes force to include freshness value of content while deciding about caching the content \cite{QuevedoCorujoAguiar2014a}-\cite{QuevedoCorujoAguiar2014}-\cite{HailAmadeoMolinaroEtAl2015}. While content popularity has been included in caching decision in \cite{VuralNavaratnamWangEtAl2014} but still there is a need to explore the popularity of content using a simple method.

A lot of research has been conducted for caching placement strategies while most of the research efforts suggest LRU as appropriate cache replacement strategy \cite{BaccelliMehlisHahmEtAl2014}-\cite{HailAmadeoMolinaroEtAl2015}-\cite{HailAmadeoMolinaroEtAl2015a}-\cite{zhang2017user}-\cite{fan2017caching}. The work in \cite{hahm2017low} designs and thoroughly analyses a cooperative caching scheme that maximizes sleeping cycles and minimizes energy consumption of constrained IoT nodes. They show in theory and experiment that a smart replication strategy can indeed save significant resources while increasing the content availability throughout a wireless IoT system. Cache coherency protocols are almost entirely missing from current literature and hold a lot of potential to be explored for IoTs.

Above all, a complete caching management system is still not present in the current literature. Caching management system should address the responsibilities of IoT nodes about sharing constraints to ensure privacy and security of IoT applications and about the validity of contents in a node.    
\subsection{Content Routing and Information /Content Delivery}
ICN-IoTs involves data routing and forwarding mechanisms when consumer node is far-away from producer node or indirectly connected in the multi-hop fashion. Mostly ICN architectures support content naming while some research efforts in ICN-IoTs support naming IoT devices \cite{AbidySaadallahyLahmadiEtAl2014}. To provide routing for these two different types of names, either content name can be directly used in routing or device name can be resolved through Name Resolution System (NRS) to find requested content \cite{icn-ietdraft02}.
\subsection{Mobility}
We refer mobility to both producer and consumer mobile nodes. Most of the ICN architecture designs argue that consumer mobility is \sob{inherently} supported while producer mobility is not completely specified. ICN mobile data consumer simply re-issue interest message and network forwards this interest towards nearest and reliable data provider or data cached node. However, for ICN-IoTs most of the nodes can act as providers/producers of information. In IoT applications like VANETs, vehicles act as information producer about the road condition, for instance, information about the accident, road construction, and can even operate as information provider when these vehicles cache data to forward to other vehicles nodes. 
Producer mobility \cite{jiang2012content} \sob{categorization} is provided in \cite{zhang2016survey}, these four approaches (tracing and mapping mobile producer, data can be moved to a near stationary place or data can be regenerated from other mobile producers in that region) can be implemented for IoT scenarios. Also a proactive technique \cite{ko2014proactive} can be \sob{investigated} for IoTs \sob{environment}. To cope with provider mobility in ICN, an initial draft is presented in \cite{auge2015anchor} through simple and easy to maintain anchor-less approach. We argue that this approach should be explored and can become very beneficial in IoT constraint-oriented devices having limited resources.
\subsection{Privacy and Security}
A full of the potential research area is privacy and security of both user requests and data in ICN-IoTs applications. Although ICN provides authentication and access control at content level but content requests are stored in ICN intermediate routers and can be tracked by attackers \cite{ndnsurvey2016}. Thus to maintain privacy at the router level between user and producer, privacy algorithms are required. Also, it is still not standardized to decide whether intermediate routers will be present in ICN-IoTs applications or not \cite{lindgren2016design}. 
Moreover, public key infrastructure (PKI) is very complex to implement for constraint oriented devices as it requires much power in the implementation of trust management and key generation \cite{ShangDingMarianantoniEtAl2014}-\cite{burke2014secure}. Therefore, light cryptography and light hash function can be evaluated and hence modified for constraint-oriented devices. Keys generation and management that \sob{include} both key revocation lists and key distribution processes are still needed to \sob{explore} further for IoTs applications. In addition, a significant research area is control access strategies in which user authentication, their corresponding access privileges, cache access, and updates are needed to be investigated for IoTs applications. \sob{Moreover}, security of sensitive information, spoofing and sniffing is highly needed to explore and address as highlighted in \cite{AbdAllahHassaneinZulkernine2015}. In \cite{saxena2017design} ICN-based safety is discussed in healthcare applications and can be explored for other IoT applications like smart home, smart grid and smart traffic.

In a nutshell, a complete mechanism ensuring both privacy and security for IoT data and applications is missing in current literature and therefore there is a strong need to design a holistic solution in this perspective. 
\subsection{Edge Computing (In-network Computation) and Cloud Computing}
From IoTs perspective, in-network computation is a mechanism through which data collected from constraint-oriented sensors initially processed and later on, refined data is transmitted towards the requested host. In-network computation is necessary to reduce the amount of produced data while lessening storage and high processing requirements. Other advantages of in-network computation include easy management of mobile nodes, less and refined cached data, simple data routing and forwarding and hence it can improve network-life, battery-life at the cost of simple and optimal in-network computation algorithms. In-network computation is the base for a new trend known as edge computing. As we mentioned earlier in Table~\ref{table-IoT-LifeCycle-References} and Fig.~\ref{iot_Life_Cycle} that cloud computing is the main force which is involved in IoT life cycle to process and manage IoT contents. As cloud computing separates producer and consumer of information, which increases delay and bandwidth during the transmission and reception of information to central servers of cloud computing just for processing of data and management of information. Moreover, it poses many privacy concerns which can occur during the reception and transmission of content to/from consumer/producer. Due to these disadvantages, a new paradigm with the name fog computing is introduced to shift computing and storage capabilities towards the end node or edge node of the network. Due to the involvement of edge nodes and edge routers, fog computing is also known as edge computing \cite{chiang2016fog}. As edge computing need to cache data before its processing and in ICN-IoT, ICN enables IoT devices to cache data naturally. Thus in ICN-IoT caching with edge computing, IoT devices can also process the cached data.

 Moreover, in ICN-IoT, it is encouraged to cache data near to end consumers (end nodes) which helps edge computing further. As a consequence, edge computing (in-network computation) becomes a key player for ICN-IoT caching. In IoT applications like virtual and augmented reality based games which require real-time behavior with almost zero-delay can benefit from edge computing \cite{premsankar2018edge}. A distributed edge computing mechanism divides the whole task among different devices of the network and ICN instance name function networking (NFN) can improve the working of many ICN-IoT applications including smart-home and health, VANETs and smart grid \cite{sifalakis2014information}. This NFN further explored for IoTs and extended with scheduling algorithm \cite{wang2016cs}. Three resolution strategies are defined to support edge find or execute (EdgeFoX), Find-and-Execute (FaX) and Find-or-Pull-and-Execute (FoP)aX. These strategies can be applied to a smart home or smart building \cite{scherb2018resolution}. Further, roles and addition of added nodes to perform the in-network computation is needed to explore. Moreover, there is a need to explore that how in-network computation will be performed in case of mobile nodes with and without caching.

Another way to perform ICN-IoT data processing and computation by employing cloud computing \cite{ravindran2013towards}. Clouds can share the burden of processing while providing high storage and can be used for calculating the analytics of any specific ICN-IoT application. For instance, high electricity usage can be calculated and can be seen in any specific town of the city. Therefore, cloud-assisted ICN-IoTs are needed to design that can, perform complex calculations, provide big storage and act as the backup in case of mobile devices \cite{borgia2016mobile}.


\subsection{Content Discovery}
In ICN, produced content is published by the producer by placing the corresponding name in the nearest ICN-based router and it is stored in the router to fulfil further consumer queries. In ICN-IoTs, consumer requests can be satisfied in two ways: (i) content is provided from the nearest router, (ii) content is fetched \sob{directly} from the content producer. While in second case, consumer devices may need data with specific constraints like freshness \cite{QuevedoCorujoAguiar2014a}-\cite{HailAmadeoMolinaroEtAl2015a}. To provide content accessibility in efficient way through ICN, packet formats must be specified and re-designed to cope such needs that could lead to easy content discovery and efficient delivery towards the consumer. \textit{Interest Message} and \textit{Data Message} should be modified in order to support push-type communication in ICN-IoTs \cite{AmadeoCampoloMolinaro2014}. For this, name-based aggregation can provide improved latency and efficient information lookup \cite{AbidySaadallahyLahmadiEtAl2014}. However, issues related to content discovery include the need to resolve: (i) How to name continuously produced contents to provide efficient look-up? (ii) How to manage content discovery efficiently in highly dynamic environments like VANETs? and (iii) How to map and search contents from named-devices corresponding to content requests efficiently?. 
\subsection{Quality of Service (QoS)}
As ICN-IoTs have to drive highly heterogeneous and constraint-oriented devices, e.g., limited memory, limited battery life and specific processing unit. With these constraint-oriented devices, ICN-IoTs specific applications QoS needs, e.g., low latency for VANETs, smart city and smart grid, better scalability and high reliability for smart health, smart grid, smart house and smart personal applications, should be satisfied and are not yet considered to be explored. Therefore, there is an urgent need to design QoS-aware protocols to evaluate the performance of ICN-IoTs for latency, reliability, resource-consumption and scalability.
ICN has much potential to improve delay and save bandwidth to satisfy different QoS requirements. ICN striking features in-network caching, any-cast, multi-cast, adaptability to mobile devices and dynamic environments and content security at the network layer reduces many efforts that need to be done with TCP/IP.  
\subsection{Business Strategies and Models}
It is essential as well as critical to design business models for ICN-based IoTs because IoTs is known to be very advantageous and useful in our daily life. Therefore, business-strategy-makers are highly invited to put efforts to decide policies for ICN-based IoTs. 

We identify some main questions that are needed to be explored and answered by the research community from the perspective of major entities involved in the designing of these strategies. From the consumer side, researchers need to investigate following questions: \textit{What benefits will customers receive by sharing the data of their own servers, let's say, data from home server, to be cached?}, \textit{How will the privacy of a consumer be endured?} and \textit{How much a consumer have to pay to upgrade to ICN-based IoTs solutions?}. Potential solutions for this can include, for instance, to provide quality data through caching, smart-home owners can get some extra free electricity or extra coaching to reduce their bills, smart-car-owners can avail free driving tips or road condition notifications in advance. From service-providers one need to look for these following questions: \textit{How ICN-based IoTs will help to improve the QoS?}, \textit{How it will assist to increase revenue growth?} and \textit{What they would need to offer customers for caching the data?}. 
Most importantly, every country government needs to participate in deciding the extent of data sharing. 

However, we are far beyond from this phase of designing business models and therefore, business policymakers need to involve stakeholders, consumers and manufacturers to decide analytical consensus.  
\section{Conclusions}
We discussed and presented related literature of both new paradigms IoTs and ICN. 
Then, requirements and challenges to build a reliable and inter-operable communication network architecture for IoTs are presented. Through this paper, we have also discussed ICN suitable features, different ICN projects for the future Internet design and their resulting ICN-based network architectures for IoTs. ICN projects are briefly summarized in terms of their corresponding feasibility for IoTs in terms of naming schemes, caching mechanisms, security and mobility support. Mapping of IoTs communication network architecture requirements against ICN striking and supporting features is presented. Furthermore, we discussed ICN-based solutions/architectures for IoTs to present the applicability of ICN for IoTs. Then, we presented and classified ICN-IoT state-of-the-art literature into four categories of naming, caching, security and mobility, and presented in four different sections. 
Moreover, compatible operating systems and simulators for ICN-based IoTs are discussed in the next section. In the end, we present identified research gaps which need research community attention to build ICN-based network architecture for IoTs.
\section*{\sobia{Acknowledgment}}
This research is supported by the Computer Engineering Department (CPED) of the University of Engineering and Technology (UET), Taxila, Pakistan under a Full-time research scholarship, and in close collaboration with both University of West London, UK and Waterford Institute of Technology, Ireland. 
\bibliographystyle{IEEEtran}
\bibliography{IEEEabrv,iot6}

\end{document}